\documentclass[prd,tightenlines,floatfix,showpacs,preprintnumbers,nofootinbib,eqsecnum,superscriptaddress,preprint]{revtex4-1}

\usepackage{epsfig}
 
\def\d{{\rm d}}

\def\le{\left[ }
\def\re{\right] }

\def\beq{\begin{equation}}
\def\eeq{\end{equation}}
\def\bea{\begin{eqnarray}}
\def\eea{\end{eqnarray}}

\begin{document}

\preprint{MS-TP-16-29}

\title{Nuclear parton density functions from jet production \\ in DIS at the EIC}

\author{M.\ Klasen}
\email{michael.klasen@uni-muenster.de}
\affiliation{Institut f\"ur Theoretische Physik, Westf\"alische
 Wilhelms-Universit\"at M\"unster, Wilhelm-Klemm-Stra{\ss}e 9,
 D-48149 M\"unster, Germany}

\author{K.\ Kova\v{r}\'ik}
\email{karol.kovarik@uni-muenster.de}
\affiliation{Institut f\"ur Theoretische Physik, Westf\"alische
 Wilhelms-Universit\"at M\"unster, Wilhelm-Klemm-Stra{\ss}e 9,
 D-48149 M\"unster, Germany}

\author{J.\ Potthoff}
\email{j\_pott04@uni-muenster.de}
\affiliation{Institut f\"ur Theoretische Physik, Westf\"alische
 Wilhelms-Universit\"at M\"unster, Wilhelm-Klemm-Stra{\ss}e 9,
 D-48149 M\"unster, Germany}

\date{\today}


\begin{abstract}
We investigate the potential of inclusive-jet production in deep-inelastic
scattering (DIS) at a future Electron-Ion Collider (EIC) to improve our
current knowledge of nuclear parton density functions (PDFs). We demonstrate
that the kinematic reach is extended similarly to inclusive DIS, but that the
uncertainty of the nuclear PDFs, in particular of the gluon density at low
Bjorken-$x$, is considerably reduced, by up to an order of magnitude
compared to the present situation. Using an approximate
next-to-next-to-leading order
(aNNLO) calculation implemented in the program {\tt JetViP}, we also make
predictions for three different EIC designs and for four different light and
heavy nuclei.
\end{abstract}


\pacs{
12.38.Bx, 
13.60.Hb, 
13.87.Ce, 
24.85.+p 
}

\maketitle


\section{Introduction}

Parton density functions (PDFs) are important fundamental quantities
describing our current knowledge of (primarily longitudinal) momentum
distributions of quarks and gluons in protons ($p$) and nuclei ($A$).
As intrinsically non-perturbative quantities, they are usually extracted
by fitting perturbatively calculated cross sections, in particular of inclusive
Deep-Inelastic Scattering (DIS) in $ep$ or $eA$ collisions and the Drell-Yan (DY)
process in $pp$ or $pA$ collisions, to experimental data. The predictive
power of this procedure lies in the universality, i.e.\ the process-independence
of the PDFs guaranteed by the QCD factorization theorem \cite{Collins:1989gx}
and their perturbative evolution with the resolution scale $Q^2$, which is
typically the virtuality of the exchanged vector boson. For proton PDFs, the
DESY
HERA $ep$ collider delivered an unprecedented wealth of data, which now allows
for precise theoretical predictions of CERN LHC cross sections, required
notably for the searches for physics beyond the Standard Model
\cite{Butterworth:2015oua}.

In contrast, nuclear PDFs lag considerably behind. DIS (and DY) data so far
only exist in fixed-target kinematics, which considerably restricts their
range in Bjorken-$x$ and $Q^2$, and only with limited statistics for
various nuclei
(typically He, C, Ca, Fe, W, Au and Pb), often as ratios to $e$D or $pp$
data. A future Electron Ion Collider (EIC) would therefore have a strong
impact, in particular on understanding the small- and large-$x$ regions of
nuclear shadowing and the EMC effect, respectively, and on pinning down the
poorly restricted gluon densities in nuclei, as laid out in detail in the EIC
White Paper \cite{Accardi:2012qut} and also discussed at the recent POETIC 7
conference \cite{armesto}. Current analyses of nuclear PDFs like DSSZ
\cite{deFlorian:2011fp}
and nCTEQ15 \cite{Kovarik:2015cma} have mostly relied on inclusive pion data
from BNL RHIC to restrict the nuclear gluon PDF with the disadvantage that
these data depend also on the pion fragmentation function, which may
furthermore be modified by medium effects \cite{deFlorian:2011fp}. The
importance of nuclear PDFs thus also lies in the fact that their knowledge is
mandatory for a clean separation of cold and hot nuclear effects in the
determination of the properties of the Quark Gluon Plasma. In addition
to inclusive pion data in D-Au collisions at BNL RHIC, the EPPS16
update to the EPS09 analysis uses also LHC $p$-Pb data on inclusive dijet
production \cite{Eskola:2009uj,Armesto:2015lrg}, while the update of the
HKN07 analysis has focused on neutrino data \cite{Hirai:2007sx} and the
question of universality of neutral and charged current DIS
\cite{Schienbein:2009kk}. Vector boson and (slightly virtual) photon
production have also been suggested \cite{Brandt:2013hoa} and employed
\cite{Kusina:2016fxy} as possible improvements.

In this paper, we study the impact of inclusive jet measurements in DIS at
a future EIC on the determination of nuclear PDFs. In contrast to inclusive
DIS, jet production is not dominated by quark scattering, but also
sensitive to gluon-initiated processes. At the same time, only cold nuclear
effects are measured in (pointlike) electron-ion collisions in contrast
to $AA$ collisions and possibly even $pA$ collsions, where collective effects
are currently hotly debated \cite{Abelev:2012ola}. Our calculations are
based on previous work on jet production in photoproduction
\cite{Klasen:1995ab} and DIS \cite{Klasen:1997jm} at next-to-leading order
(NLO), which we have recently systematically improved in both cases to
approximate next-to-next-to-leading order (aNNLO) \cite{Klasen:2013cba,%
Biekotter:2015nra} with a unified approach to soft and virtual corrections
\cite{Kidonakis:2003tx}. Note that very recently also full NNLO calculations
of inclusive jet \cite{Abelof:2016pby} and dijet production
\cite{Currie:2016ytq} in DIS have become available, which show that the NNLO
corrections are moderate in size, except at the kinematical edges, and that
their inclusion leads to a substantial reduction of the scale variation
uncertainty on the predictions. We emphasize again that our focus here is not
the impact of higher-order corrections in $ep$ collisions, but rather the
sensitivity of this process to nuclear effects in $e$A collisions at the EIC.

The remainder of this paper is organized as follows: In Sec.\ \ref{sec:2},
we describe our theoretical setup, including in particular our choices of
renormalization and factorization scales and PDF sets. In Sec.\ \ref{sec:3},
we review the proposed experimental conditions for the two possible EIC
designs and their detectors, i.e.\  BNL's proposal to add an electron ring to
the existing Relativistic Heavy Ion Collider (eRHIC) and Jefferson
Laboratory's proposal to build a Medium energy Electron Ion Collider (MEIC)
or Jefferson Laboratory EIC (JLEIC) using the upgraded 12 GeV Continuous
Electron Beam Accelerator Facility
(CEBAF). We base our assumptions on publicly available information from
the EIC White Paper \cite{Accardi:2012qut} and updates shown at the POETIC 7
conference \cite{yoshida,mueller}. Our numerical results are
presented in Sec.\ \ref{sec:4} for a variety of EIC realizations and nuclei
of different mass. Here, we also quantify the effect of higher-order
corrections and, more importantly, estimate the impact of a future EIC on the
reduction of nuclear PDF uncertainties. Our conclusions and an outlook to
further studies are given in Sec.\ \ref{sec:5}.


\section{Jet production in DIS at approximate NNLO of QCD}
\label{sec:2}

The QCD factorization theorem allows us to write the differential cross section
for inclusive jet production on a nucleus $A$,
\beq
 \d\sigma=\sum_a\int\d y\,f_{\gamma/e}(y)
 \int\d x \,f_{a/A}(x,\mu_F)\d\sigma_{\gamma a}(\alpha_s,\mu_R,\mu_F) \, ,
\eeq
as a convolution of the photon-parton cross section $\d\sigma_{\gamma a}
(\alpha_s,\mu_R,\mu_F)$ with the flux of virtual photons in the electron,
$f_{\gamma/e}(y)$, and the PDFs of partons $a$ in the nucleus $A$, $f_{a/A}
(x,\mu_F)$. The fractional energy transfer of the electron in the nuclear
rest frame is defined as $y=(p\cdot q)/ (p\cdot k)$ with $p$ and $k$
the momenta of the incoming nucleus and electron, respectively, and $q$ the
momentum of the exchanged photon with $Q^2=-q^2$. $x$ is the longitudinal
momentum fraction of the parton in the nucleus, $\mu_R$ and $\mu_F$ are the
renormalization and factorization scales, respectively, and $\alpha_s$ is
the strong coupling, in which the partonic cross section is perturbatively
expanded.

For our NLO calculations, we employ the program {\tt JetViP}
\cite{Klasen:1997jm,Potter:1999gg}, which we have recently improved to
approximate next-to-next-to-leading order (aNNLO) \cite{Biekotter:2015nra}
with a unified approach to soft and virtual corrections
\cite{Kidonakis:2003tx}. In this approach, the NLO corrections can be
expressed in terms of a master formula,
\beq
 \d\sigma_{\gamma a}^{\rm NLO}=
 \d\sigma_{\gamma a}^{\rm LO}{\alpha_s(\mu_R)\over\pi}
 \le c_3D_1(z)+c_2D_0(z)+c_1\delta(1-z)\re \, ,
\eeq
which is ordered in terms of the leading and next-to-leading
logarithms
\beq
 D_l(z)=\le{\ln^l(1-z)\over 1-z}\re_+
\eeq
at partonic threshold ($z\to1$) in pair-invariant-mass kinematics
with $l\leq2n-1$ and $n=1$ at NLO, $n=2$ at NNLO etc. The NNLO
master formula is given in Eq.\ (2.17) of Ref.\ \cite{Kidonakis:2003tx},
as are (in the section preceding this equation) the general formul\ae\
for the universal coefficients $c_i$. The process-dependent ingredients
of the NNLO master formula were extracted from our explicit NLO calculation
whereever possible \cite{Biekotter:2015nra}.

In addition to the photon virtuality $Q^2$, inclusive jet production
depends on a second hard scale, the jet transverse momentum $p_T$.
A customary choice of scales is therefore
\bea
 \mu_R^2=(Q^2+p_T^2)/2 &\ {\rm and}\ & \mu_F^2=Q^2,
 \label{eq:2.4}
\eea
where the choice of $\mu_F$ is motivated by the fact that the same
factorization scale can be used in the calculation of jet and inclusive
DIS cross sections \cite{Andreev:2014wwa}. Jets are reconstructed in the
Breit frame using the anti-$k_T$ algorithm with a distance parameter $R=1$
in the $\eta-\phi$ plane and a massless $p_T$ recombination scheme
\cite{Cacciari:2008gp}. Within experimental errors, consistent results
were obtained with the $k_T$ algorithm \cite{Ellis:1993tq} by the H1
collaboration at DESY HERA \cite{Andreev:2014wwa}.
For the nuclear PDFs and their current uncertainties, we employ the nCTEQ15
fit with with 32 error PDFs, and we estimate the impact of the inclusive pion
production data from BNL RHIC with its nCTEQ15-np variant
\cite{Kovarik:2015cma}.


\section{Experimental conditions at an Electron Ion Collider}
\label{sec:3}

The experimental conditions at a future EIC depend on the selected
site. At BNL, the existing RHIC is planned to continue accelerating nuclei
to beam energies of up to $E_A=100$ GeV per nucleon. It would have to
be supplemented by a new electron beam with energy $E_e=16$ to 21
GeV. The center-of-mass energy would then reach $\sqrt{s}=80$ to 90
GeV and the integrated annual luminosity approximately 10 fb$^{-1}$ for the
lower and a third of that value for the higher energy \cite{Accardi:2012qut}.

At Jefferson Lab, the Medium energy Electron Ion Collider (MEIC)
would be based on the upgraded Continuous Electron Beam Accelerator Facility
(CEBAF), which provides a high-luminosity electron beam of $E_e=12$ GeV.
It would have to be supplemented by an ion accelerator that could reach
energies of $E_A=40$ GeV per nucleon, leading to a lower center-of-mass
energy of $\sqrt{s}=45$ GeV, but a higher integrated annual luminosity of
${\cal L}=100$ fb$^{-1}$ \cite{Accardi:2012qut}.

Under all three of these conditions,
the kinematic plane in $x$ and $Q^2$ would be extended considerably, as can
be seen in Fig.\ 1.5 of Ref.\ \cite{Accardi:2012qut}, i.e.\ from $x\geq4\times
10^{-3}$ in $\nu A$ DIS and $x\geq 10^{-2}$ in $eA/\mu A$ DIS to values of
$x\geq10^{-4}$ and below in the Jefferson Laboratory and BNL designs,
respectively, while simultaneously extending the range in $Q^2$ from 
$10^2$~GeV$^2$ to $10^3$ GeV$^2$ and beyond. In this way, the experimental
information on the partonic structure of heavy nuclei would soon rival that
of protons obtained at DESY HERA.
In the following section, we will provide numerical results for each of
the three accelerator designs mentioned above. Note that upgrade options exist
for both sites, which may allow to also reach beam energies of up to $E_A=100$
GeV per nucleon with the Jefferson Laboratory EIC (JLEIC)
\cite{yoshida} and annual luminosities of up to 100 fb$^{-1}$ with eRHIC at
BNL \cite{mueller}.

For both sites, similar detector requirements have been specified.
They aim at a kinematic coverage of $Q^2>$ 1 GeV$^2$ and 0.01
$\leq y\leq$ 0.95 by using either the scattered electron or the hadronic
final state with the Jacquet-Blondel method, which has proven advantageous
at very low values of $y$ at DESY HERA. The electromagnetic calorimeter
would span the rapidity range $-4<\eta<4$
\cite{Accardi:2012qut}.
No specifications have so far been fixed for the hadronic calorimeter, so
that we assume the same coverage.
At DESY HERA, jets have been reconstructed in the Breit frame down to
transverse momenta of $p_T\geq4$ GeV \cite{Andreev:2014wwa}, which we assume
to be also possible at a future EIC.


\section{Numerical results}
\label{sec:4}

We now turn to our numerical results for inclusive jet production at the EIC.
First, we investigate the dependence of various differential cross sections on
the EIC beam energies. Next, we quantify the size of NLO and aNNLO corrections
to the LO cross sections. Third, we study the dependence of the cross sections
on the type of the colliding nucleus. Finally, our main results concern a
demonstration of the current nuclear PDF uncertainty on the inclusive jet
cross sections and the impact that a future EIC might have on their reduction.

\subsection{Inclusive jet production at different EICs}

\begin{figure}
 \epsfig{file=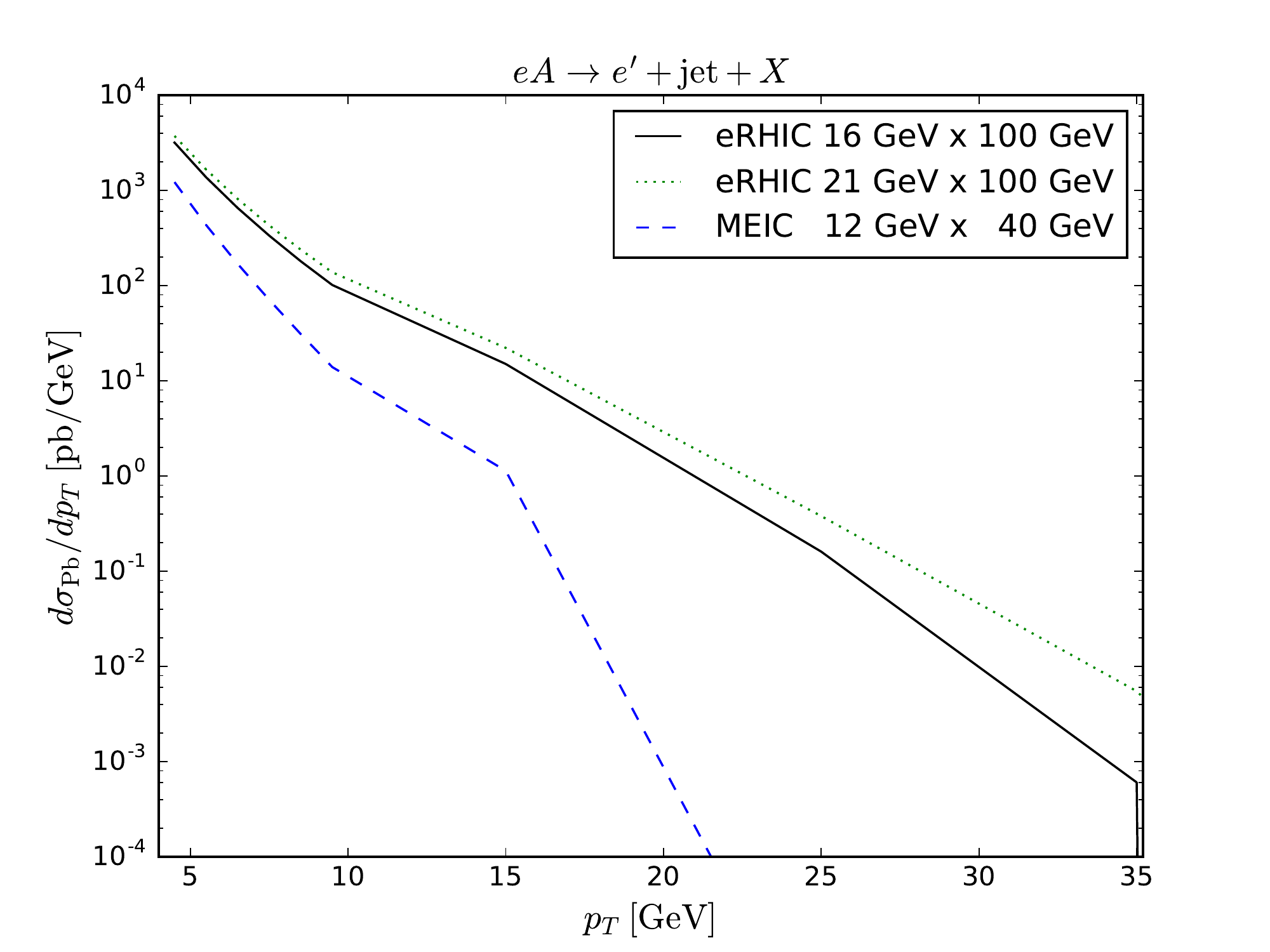,width=0.48\textwidth}
 \epsfig{file=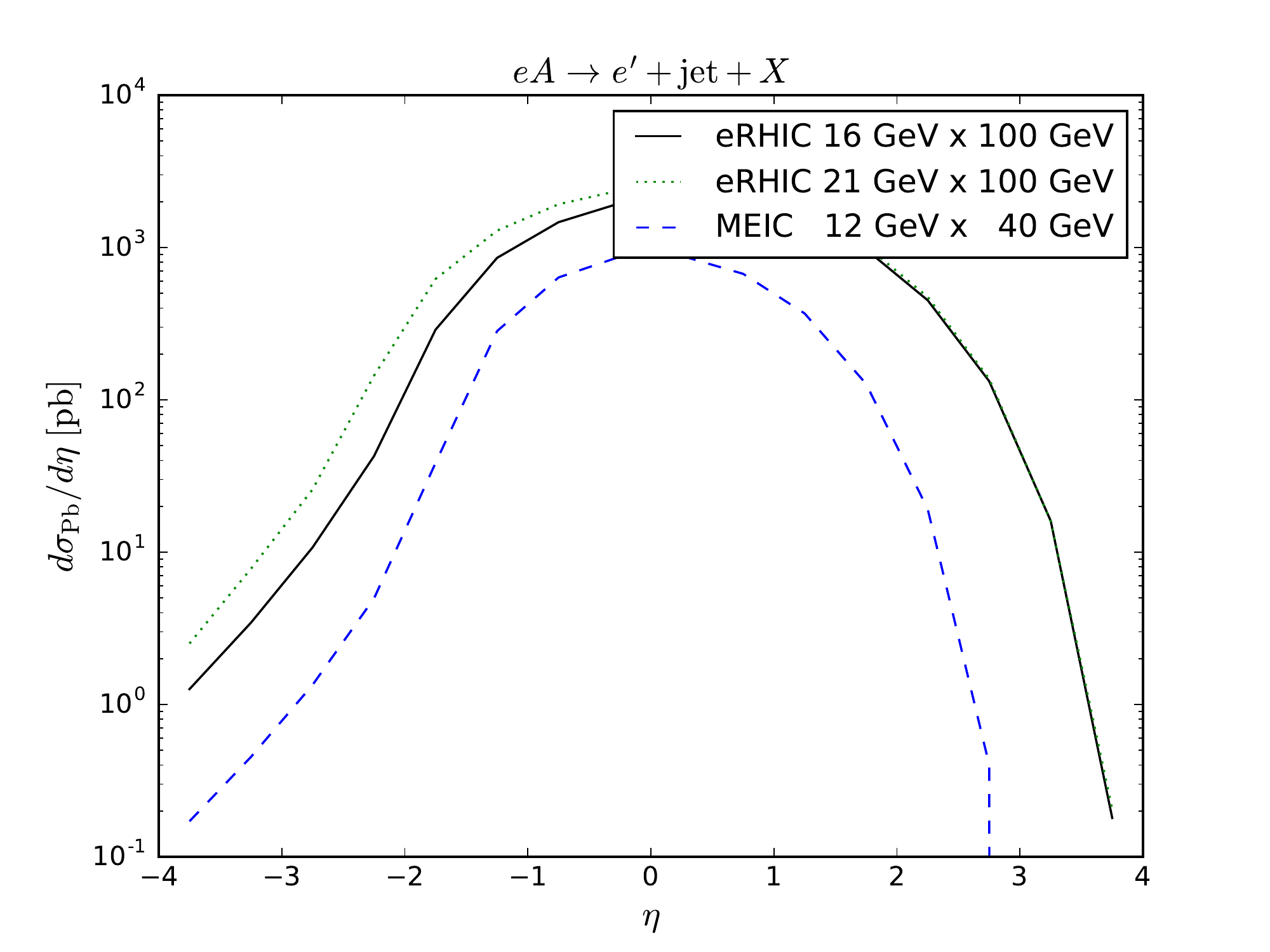,width=0.48\textwidth}
 \epsfig{file=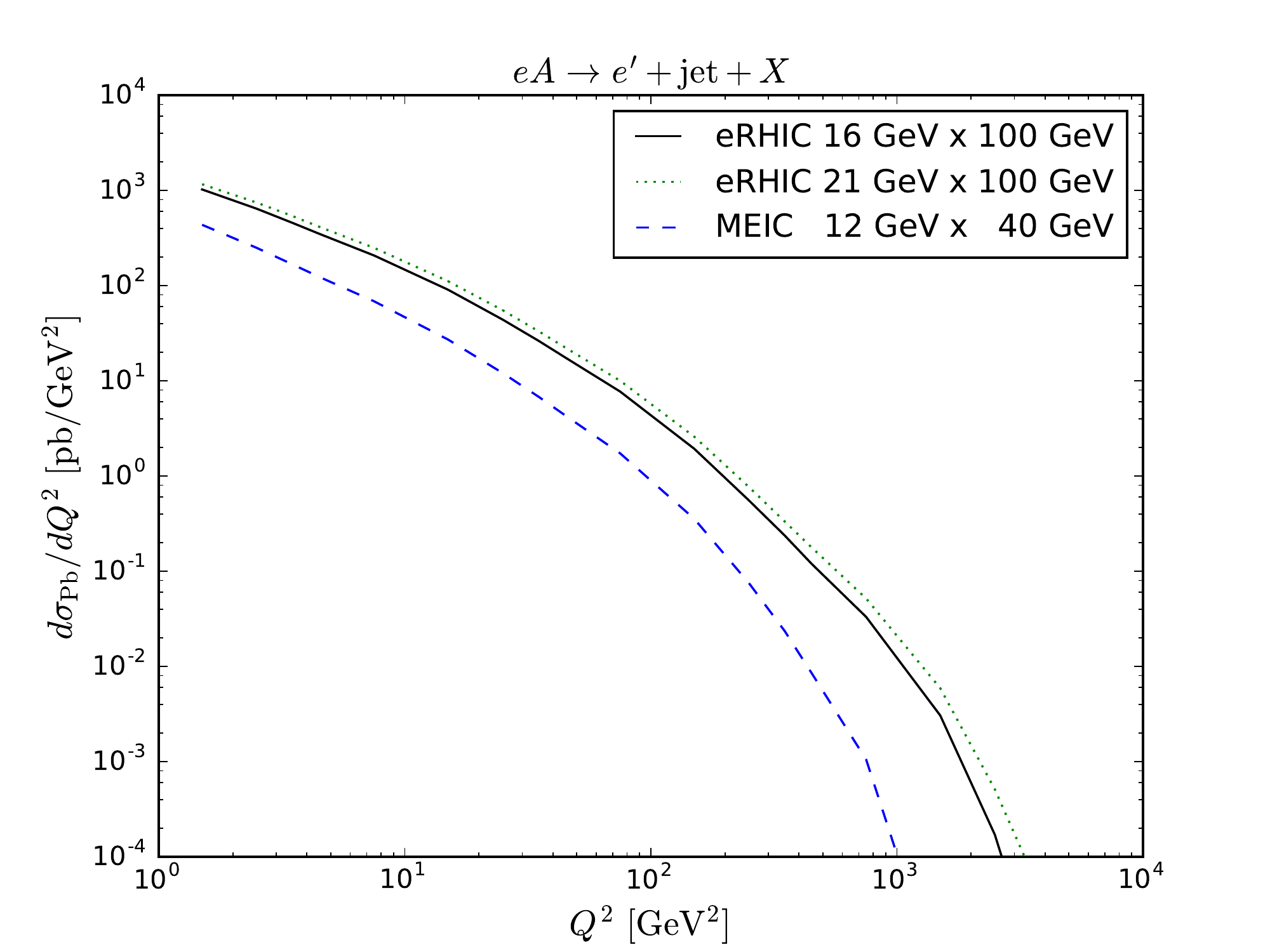,width=0.48\textwidth}
 \epsfig{file=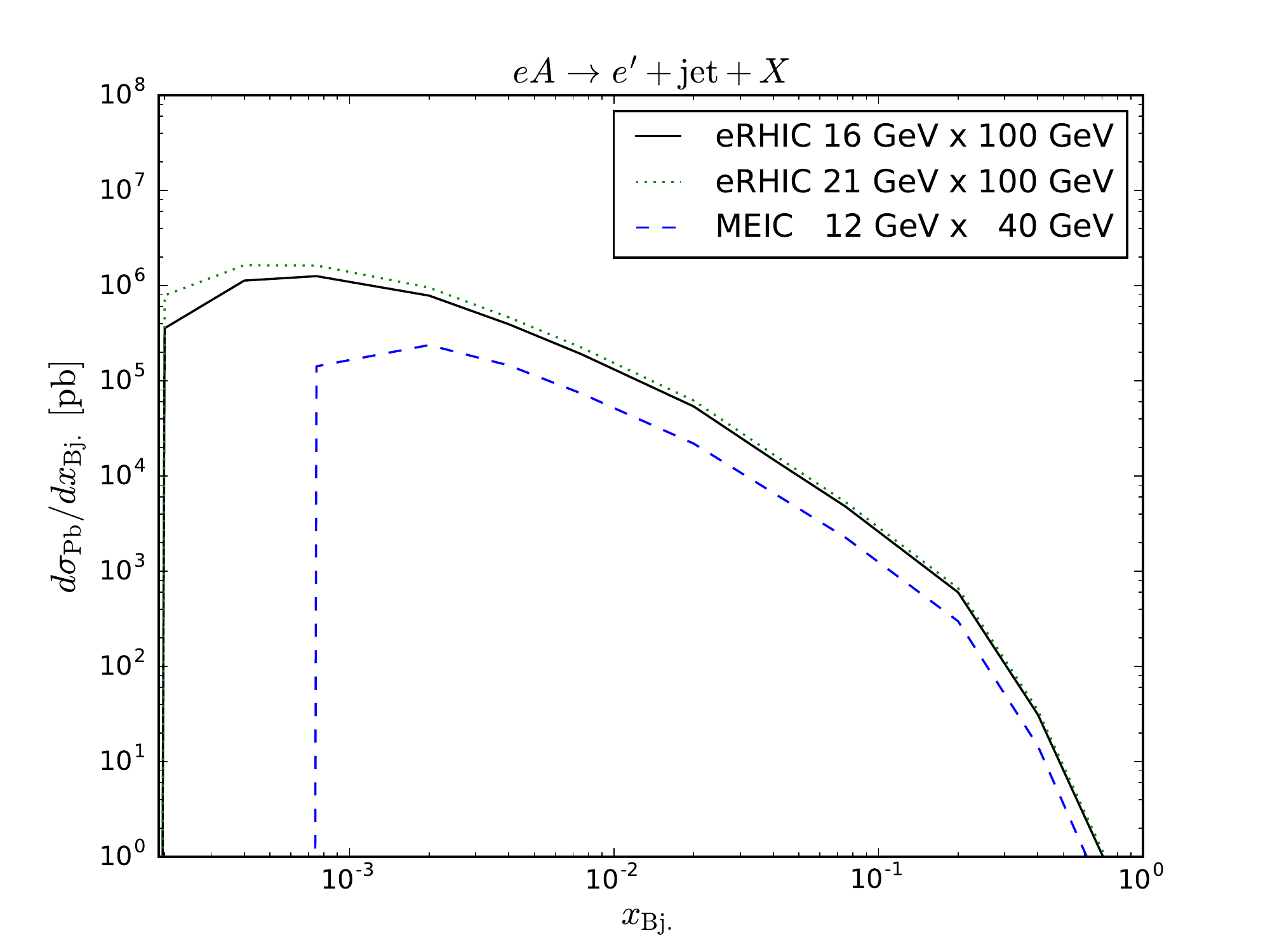,width=0.48\textwidth}
 \caption{\label{fig:1}Inclusive jet production in electron-lead ion collisions
 at eRHIC and MEIC with electron beam energies of 12 to 21 GeV and ion beam
 energies per nucleon of 40 to 100 GeV. Shown are differential cross sections
 in the jet transverse momentum (top left), rapidity (top right), photon
 virtuality (bottom left) and Bjorken-$x$ of the parton in the nucleus
 (bottom right).}
\end{figure}

In the following, we shall always display four typical differential cross
sections for inclusive jet production in DIS, i.e.\ the distributions in the
jet transverse momentum $p_T$ in the Breit frame and in the rapidity $\eta$ in
the lab frame, with the positive $z$-axis pointing in the direction of the ion
beam, as well as the DIS variables $Q^2$ and Bjorken-$x$. These four
differential cross sections are shown in Fig.\ \ref{fig:1} for $e$-Pb
collisions and three different EIC designs: the RHIC ion beam with a nominal
beam energy per nucleon of 100 GeV colliding with new electron beams of 16 GeV
(full black lines) and 21 GeV (dotted green lines), respectively, and the
CEBAF electron beam of 12 GeV energy colliding with a new ion beam of 40 GeV
energy per nucleon (dashed blue lines).

At first sight, the $p_T$-range (top left)  seems to be considerably larger in
the eRHIC designs, where it extends to 35 GeV compared to only 15--20 GeV at
the MEIC. The different nominal luminosities of 10 and 100 fb$^{-1}$ lead,
however, to a comparable number of about 100 events at 15$\pm0.5$ GeV with
only about 1 event surviving in a 20--30 GeV bin.
For the detector acceptance, we have assumed that the hadronic calorimeter
covers the rapidity range $-4<\eta<4$. The rapidity distribution in the lab
frame (top right) shows, however, that the majority of the events is contained
in the smaller range $-2<\eta<3$ at eRHIC and $-1.5<\eta<2$ at MEIC.
Similarly to the case of inclusive DIS (cf.\ again Fig.\ 1.5 of Ref.\
\cite{Accardi:2012qut}), the range in $Q^2$ would be extended in inclusive
jet production to $10^3$ GeV$^2$ at MEIC and beyond at eRHIC (bottom left),
while the range in Bjorken-$x$ extends to $10^{-3}$ and below (bottom
right). For this last distribution, which is perhaps the most interesting
for the determination of nuclear PDFs, the advantage of the eRHIC designs
with their considerably higher center-of-mass energies  over the MEIC design
is perhaps most notable.
For all four distributions, the gain in reach from a 16 to a 21 GeV electron
beam at eRHIC is, however, not very large and would probably be compensated
by the loss in luminosity.

\subsection{Inclusive jet production at LO, NLO and aNNLO}

Having established the experimental reaches in the relevant kinematic
distributions, we now turn to the more theoretical aspect of the impact
of higher-order corrections. Generally, the K-factors, i.e.\ the ratios of
the NLO or aNNLO cross sections to those at LO, are not constant, but depend
on the kinematic variables of the studied process, in particular those that
set the perturbative scales entering the strong coupling and PDFs. Large
corrections are expected at low scales and when the coupling or PDFs are
large, and vice versa.

\begin{figure}
 \epsfig{file=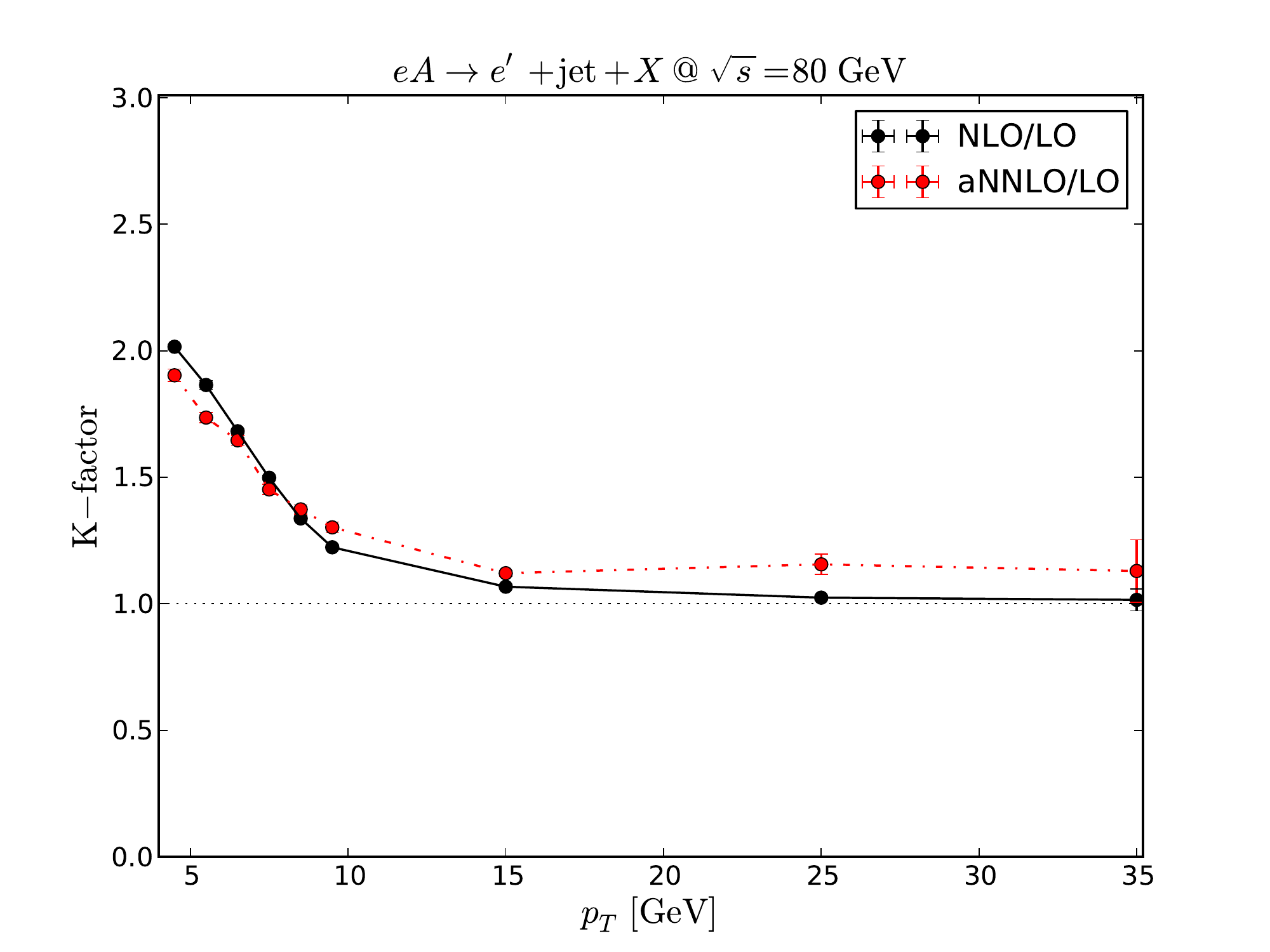,width=0.48\textwidth}
 \epsfig{file=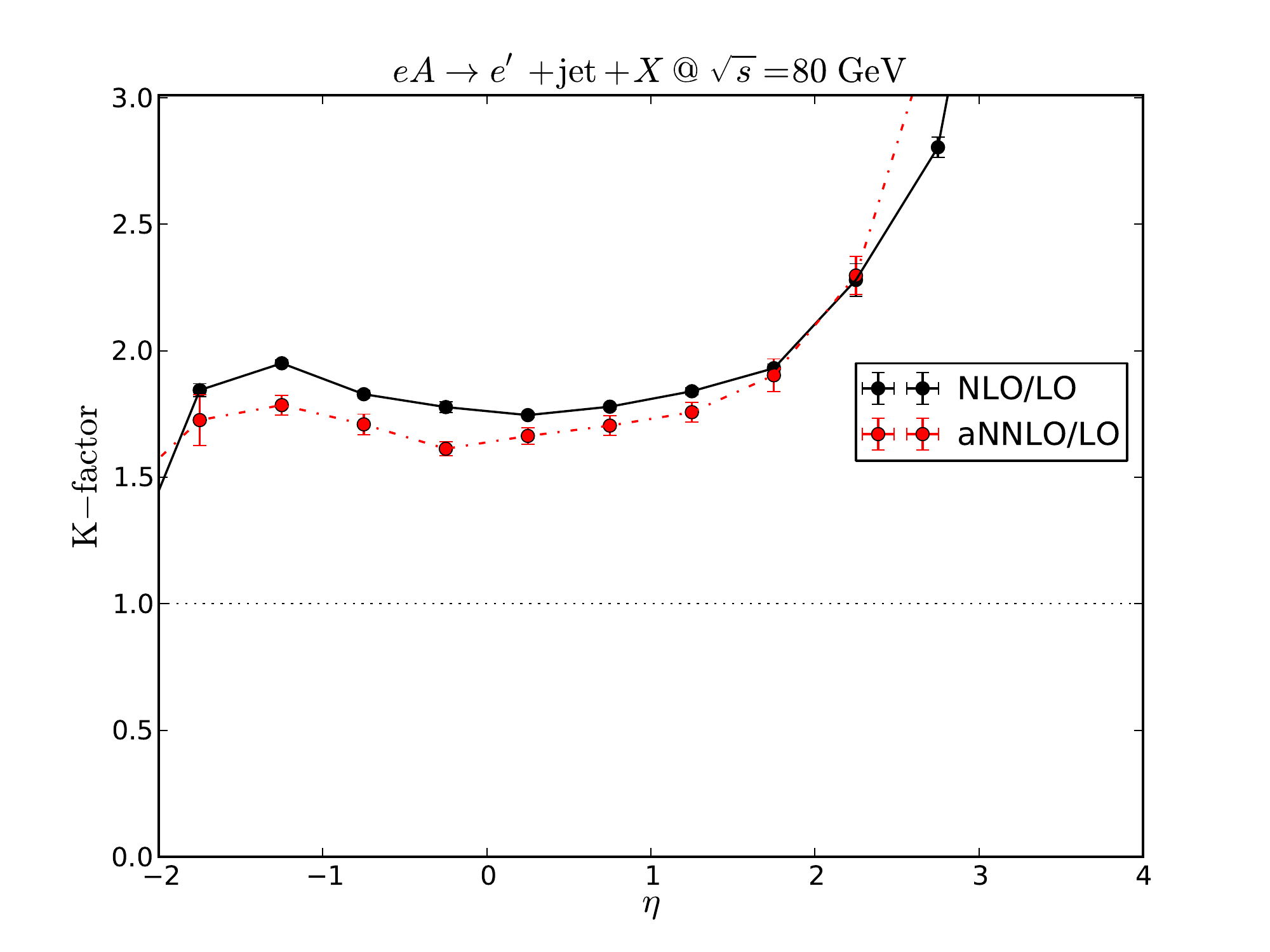,width=0.48\textwidth}
 \epsfig{file=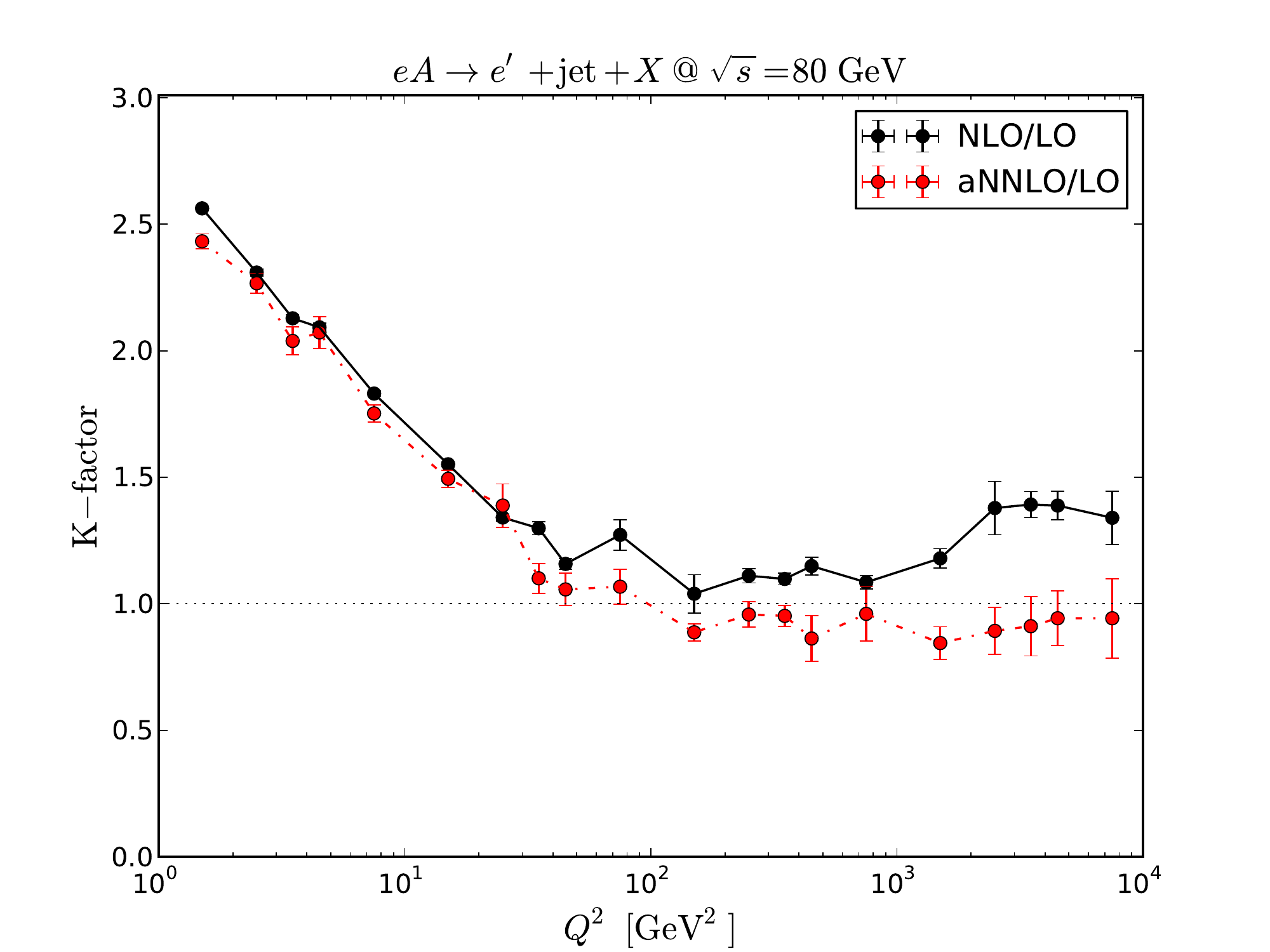,width=0.48\textwidth}
 \epsfig{file=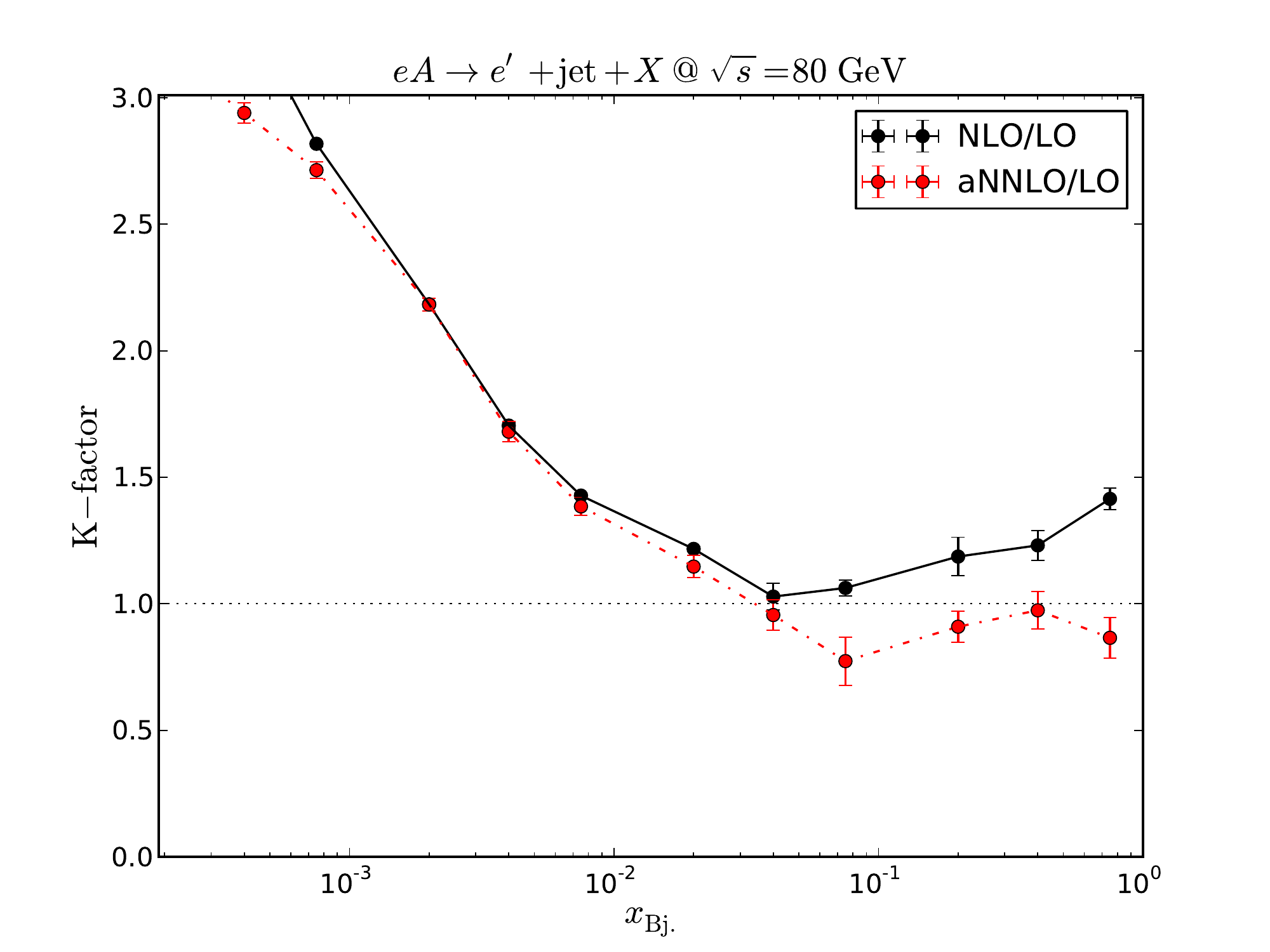,width=0.48\textwidth}
 \caption{\label{fig:2}Inclusive jet production in electron-lead ion collisions
 with beam energies of 16 and 100 GeV, respectively, at eRHIC. Shown are the
 K-factors (ratios) of NLO/LO (full black lines) and aNNLO/LO (dot-dashed red
 lines) cross
 sections as a function of the jet transverse momentum (top left), rapidity
 (top right), photon virtuality (bottom left) and Bjorken-$x$ (bottom right).
 Error bars indicate the numerical integration accuracy.}
\end{figure}

This expectation is clearly confirmed in Fig.\ \ref{fig:2}, where we show
the K-factors for the same four differential cross sections as before, but
now only for one eRHIC design with a 16 GeV electron beam and a 100 GeV
lead-ion beam. In the $p_T$-distribution (upper left) and the
$Q^2$-distribution (lower left), the NLO corrections reach a factor of
2--2.5 at low $p_T\geq4$ GeV and $Q^2\geq1$ GeV$^2$. These are the two
kinematic variables that enter the renormalization scale $\mu_R^2$, while
$Q^2$ alone sets the factorization scale $\mu_F^2$, cf.\ Eq.\ (\ref{eq:2.4}).
The corresponding cuts also set the scale in the rapidity distribution (upper
right), where the K-factor rises above 2 at forward rapidities due to the
high gluon density in this small-$x$ regime. The same rise is therefore seen
in the Bjorken-$x$ distribution (lower right) at very low values of $x$.
Since, for constant electron energy transfer, $x$ scales directly with $Q^2$,
the K-factors fall in both distributions towards higher values of these
variables.

Substantial K-factors (e.g.\ larger than 2) usually give rise to doubts about
the stability of the perturbative calculation. Often, they can, however, be
explained
by the opening-up of additional partonic channels. Here, this is in particular
the case for the splitting of low-$x$ gluons into quark-antiquark pairs which
then scatter off the virtual photon. In Fig.\ \ref{fig:2} the stability of the
perturbative calculation is also established by the fact that the aNNLO
K-factors (dot-dashed red lines) corrections are very similar to those at NLO (full
black lines). This confirms the observation in the exact NNLO calculations that the
NNLO corrections are moderate in size, but lead to a stabilization of the cross
sections with respect to variations of the renormalization and factorization
scales (see above) \cite{Abelof:2016pby,Currie:2016ytq}.

\subsection{Inclusive jet production on different nuclei}

\begin{figure}
 \epsfig{file=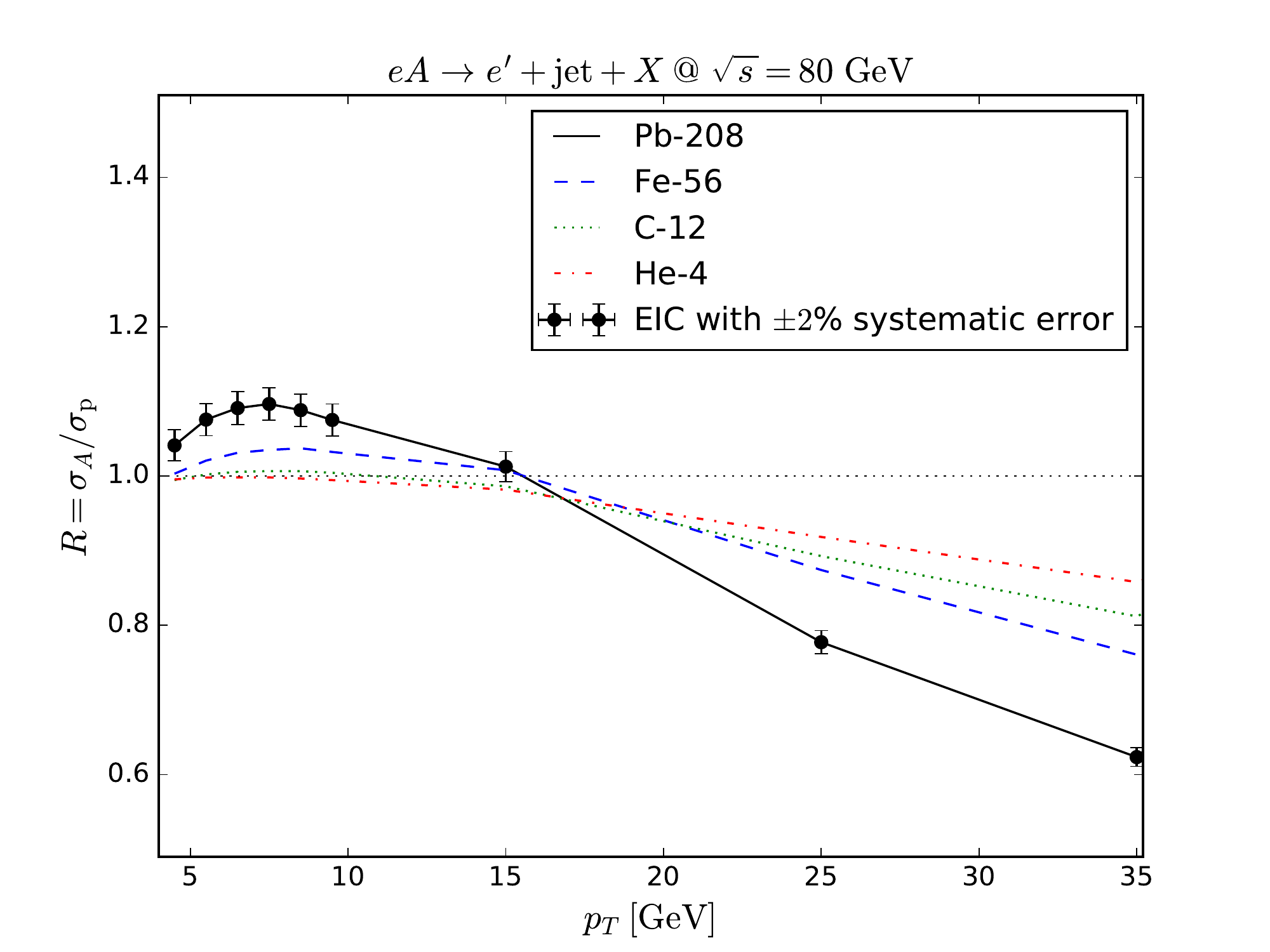,width=0.48\textwidth}
 \epsfig{file=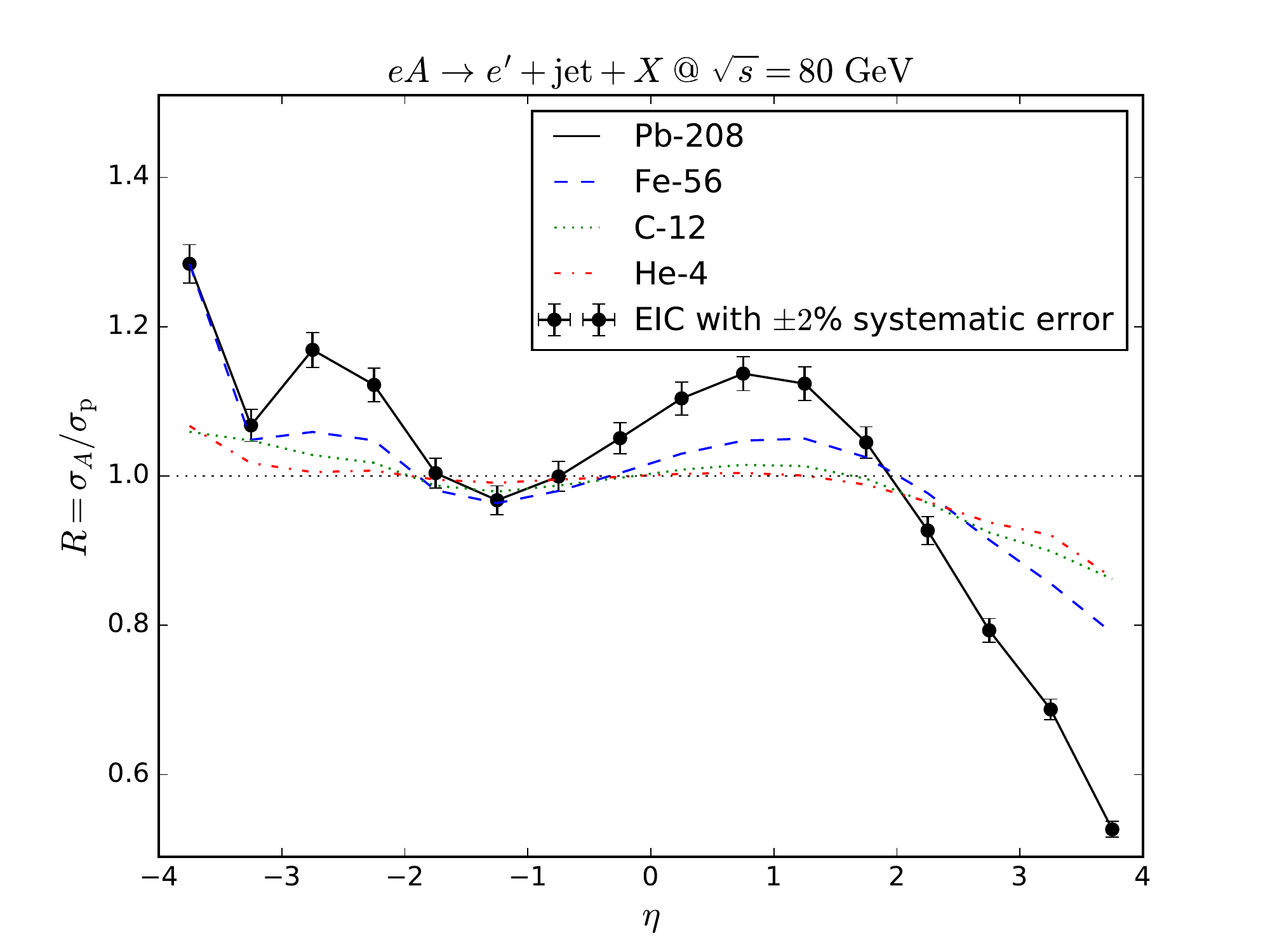,width=0.48\textwidth}
 \epsfig{file=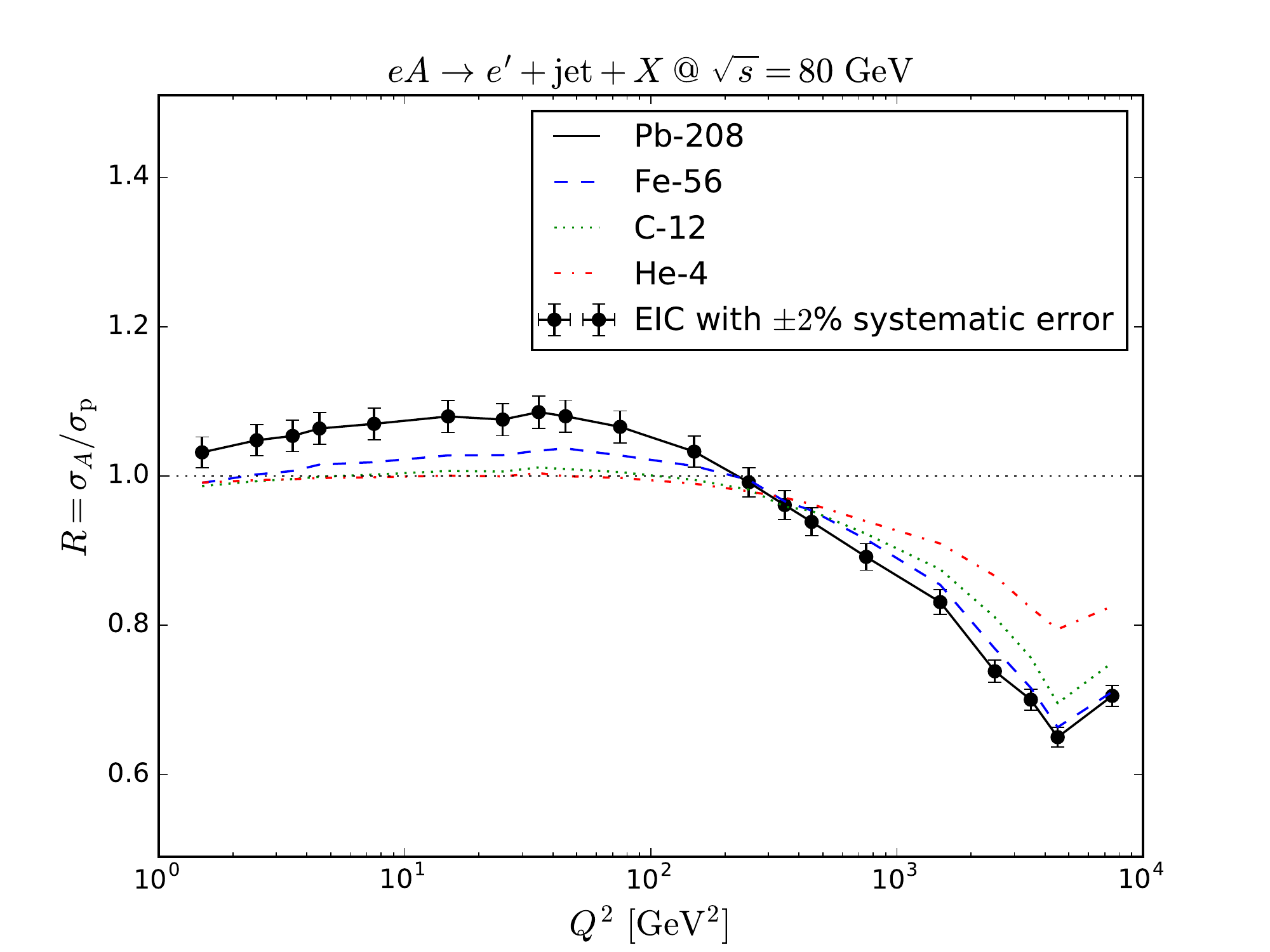,width=0.48\textwidth}
 \epsfig{file=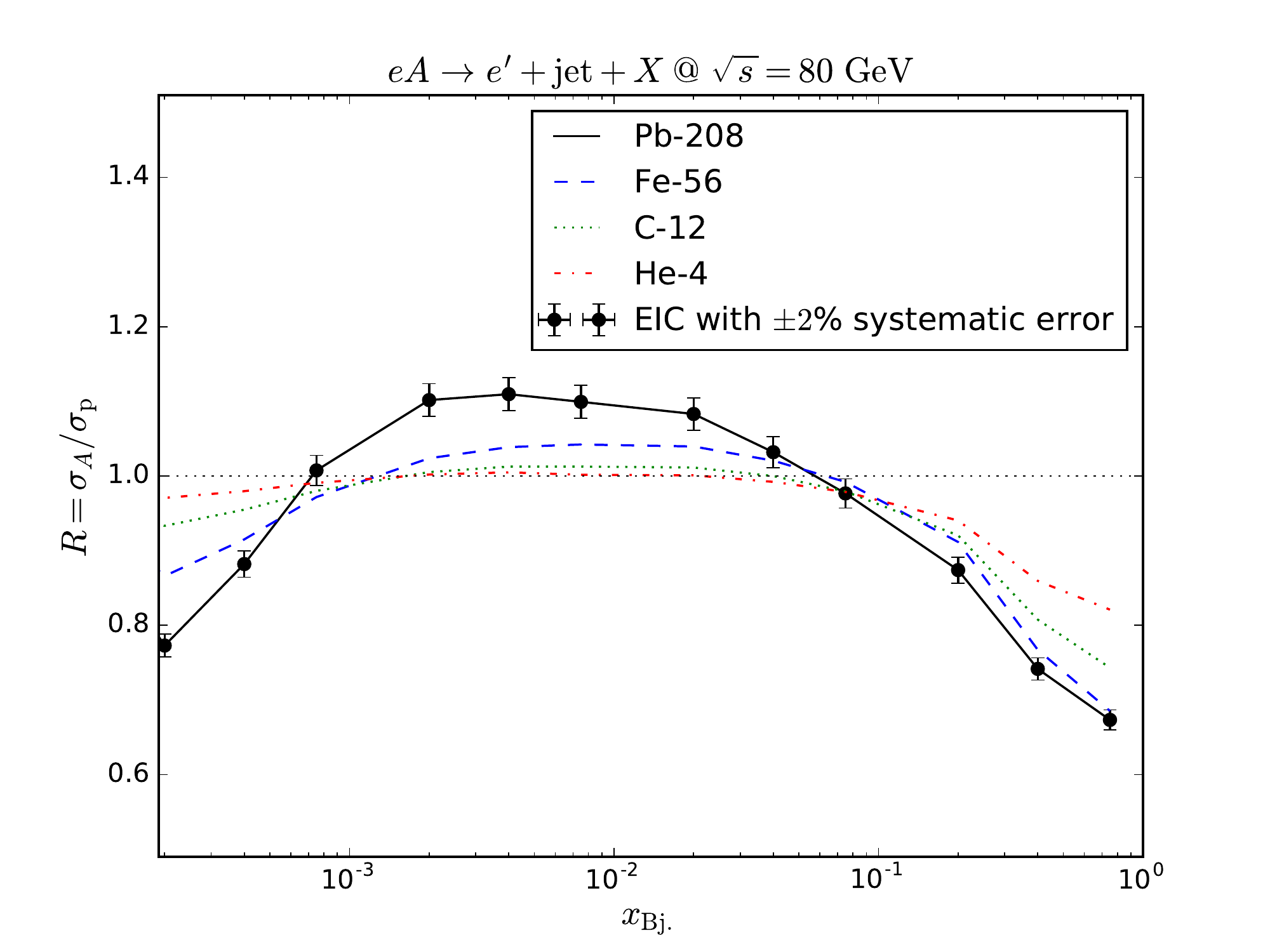,width=0.48\textwidth}
 \caption{\label{fig:3}Inclusive jet production in electron-ion collisions
 with beam energies of 16 and 100 GeV, respectively, at eRHIC for different
 nuclei: Pb-208 (full black lines), Fe-56 (dashed blue lines), C-12 (dotted green
 lines), and
 He-4 (dot-dashed red lines). Shown are the ratios of electron-ion over
 electron-proton cross sections as a function of the jet transverse momentum
 (top left), rapidity (top right), photon virtuality (bottom left) and
 Bjorken-$x$ (bottom right). Error bars indicate the expected experimental
 precision.}
\end{figure}

The main goal of our work is to demonstrate the sensitivity of an EIC to
nuclear PDF effects and to establish which regions (shadowing, antishadowing,
EMC suppression, Fermi motion) could be constrained there. We therefore
show in Fig.\ \ref{fig:3} ratios of nuclear over bare proton cross sections,
differential in the same kinematic variables as before, for typical light
and heavy nuclei: He-4 (dot-dashed red lines), C-12 (dotted green lines), Fe-56
(dashed blue lines), and Pb-208 (full black lines). The EIC design is the same as
before, i.e.\ an eRHIC machine with electron and ion beam energies (per
nucleon) of 16 and 100 GeV, respectively.

Significant reductions of 20\% and more are seen at large $p_T$ and $Q^2$,
very forward rapidities and both small and large values of $x$. The region
of small $x<10^{-3}$ with particularly high gluon and sea quark densities,
corresponding to very forward rapidities,
is known to be sensitive to nuclear shadowing induced by rescattering
\cite{Armesto:2006ph}. A particularly interesting model of nuclear shadowing
is the leading-twist approach \cite{Frankfurt:2003zd}, which is based on the
relationship between nuclear shadowing and diffraction on a nucleon and which
can be tested, among other processes, in ultraperipheral collisions at the LHC
\cite{Baltz:2007kq,Guzey:2016tek}. The shadowing effect is known to decrease
with the mass number of the nucleus \cite{Armesto:2006ph}, and this is also
clearly observed in Fig.\ \ref{fig:3}. From Pb-208 to Fe-56, C-12, and He-4,
the effect is reduced from 22\% to about 12, 6 and 3\% at $x\simeq2\times
10^{-4}$, respectively.

Reductions of up to 35\% are seen in the large-$x$ regime of the EMC
suppression, which is usually attributed to non-perturbative QCD effects on
the valence quark distributions such as multiquark clusters, dynamical
rescaling, or nuclear binding, but for which a theoretical consensus is still
missing \cite{Arneodo:1992wf,Geesaman:1995yd}. Also this reduction decreases
with the nuclear mass number, although less rapidly, i.e.\ from 35\% for
Pb-208 and Fe-56 to 25\% for C-12 and 20\% for He-4.
 
Enhancements of up to 10\% are observed at low $p_T$ and low and medium $Q^2$
as well as central rapidities and intermediate values of $x\simeq10^{-2}$.
This so-called anti-shadowing region is not only required by
momentum-conservation, but
can also be explained with constructive interference of multiple scattering
amplitudes \cite{Brodsky:2004qa,Frankfurt:2016qca}. It thus is expected to be
theoretically connected to the shadowing region, and the nuclear mass
dependence is indeed very similar. Since the experimental uncertainty on
determinations of nuclear PDFs at the EIC is expected to be dominated by a
2\% systematic error (black error bars in Fig.\ \ref{fig:3}), and not by
statistics (cf.\ Fig.\ 3.25 of Ref.\ \cite{Accardi:2012qut}), even effects
of this size should be measurable at the EIC.

\subsection{Sensitivity to nuclear parton density functions}

\begin{figure}
 \epsfig{file=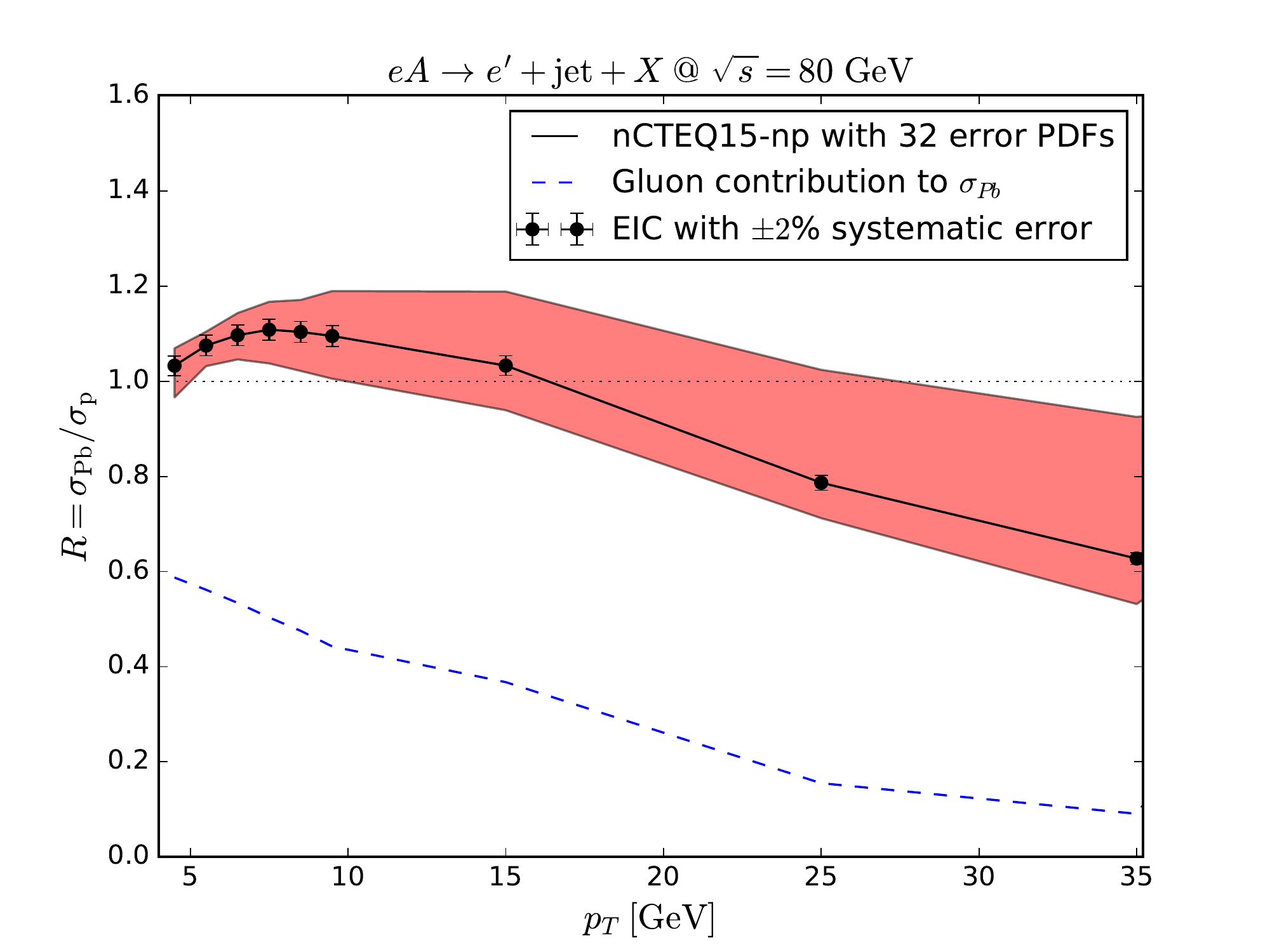,width=0.48\textwidth}
 \epsfig{file=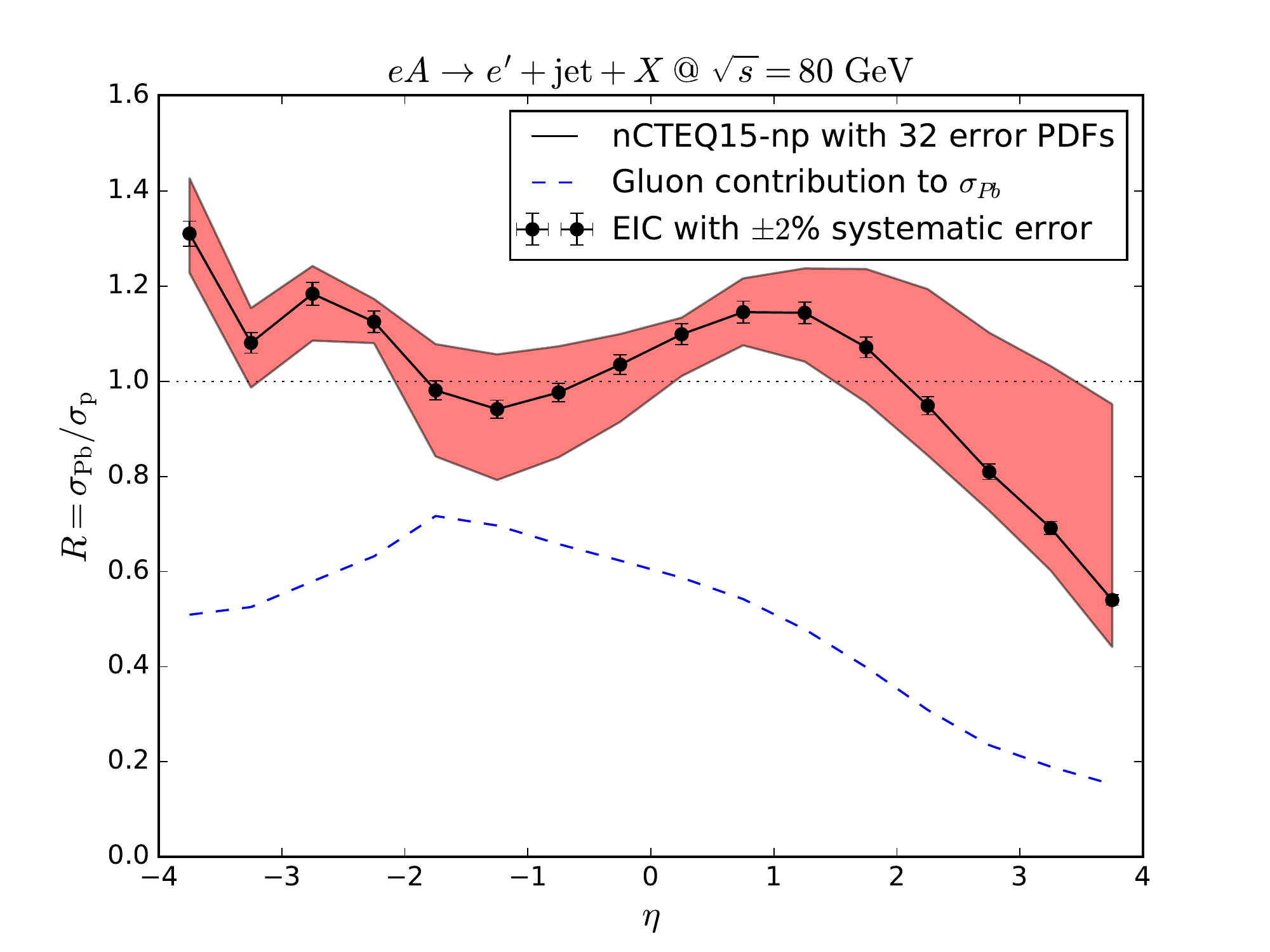,width=0.48\textwidth}
 \epsfig{file=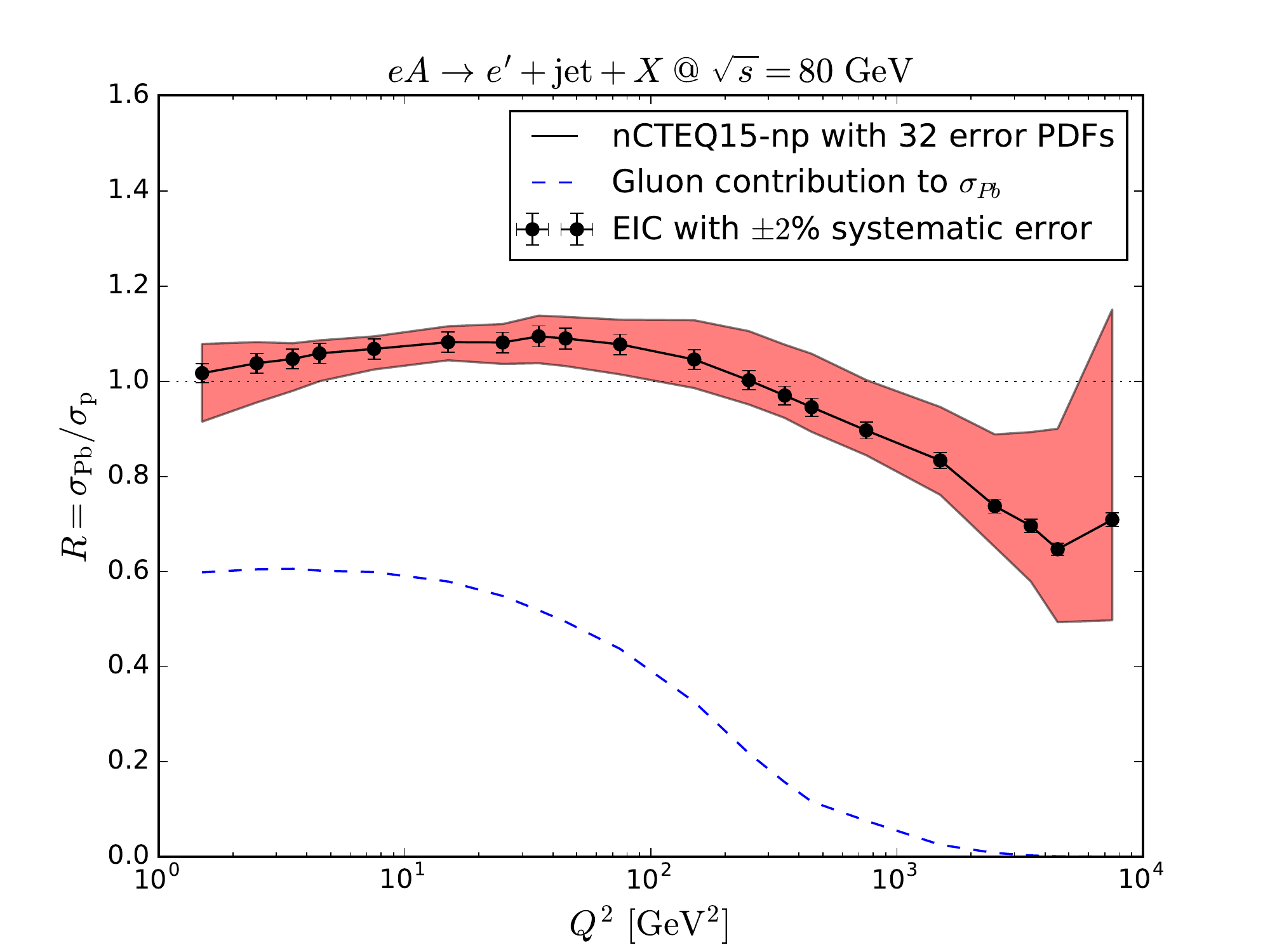,width=0.48\textwidth}
 \epsfig{file=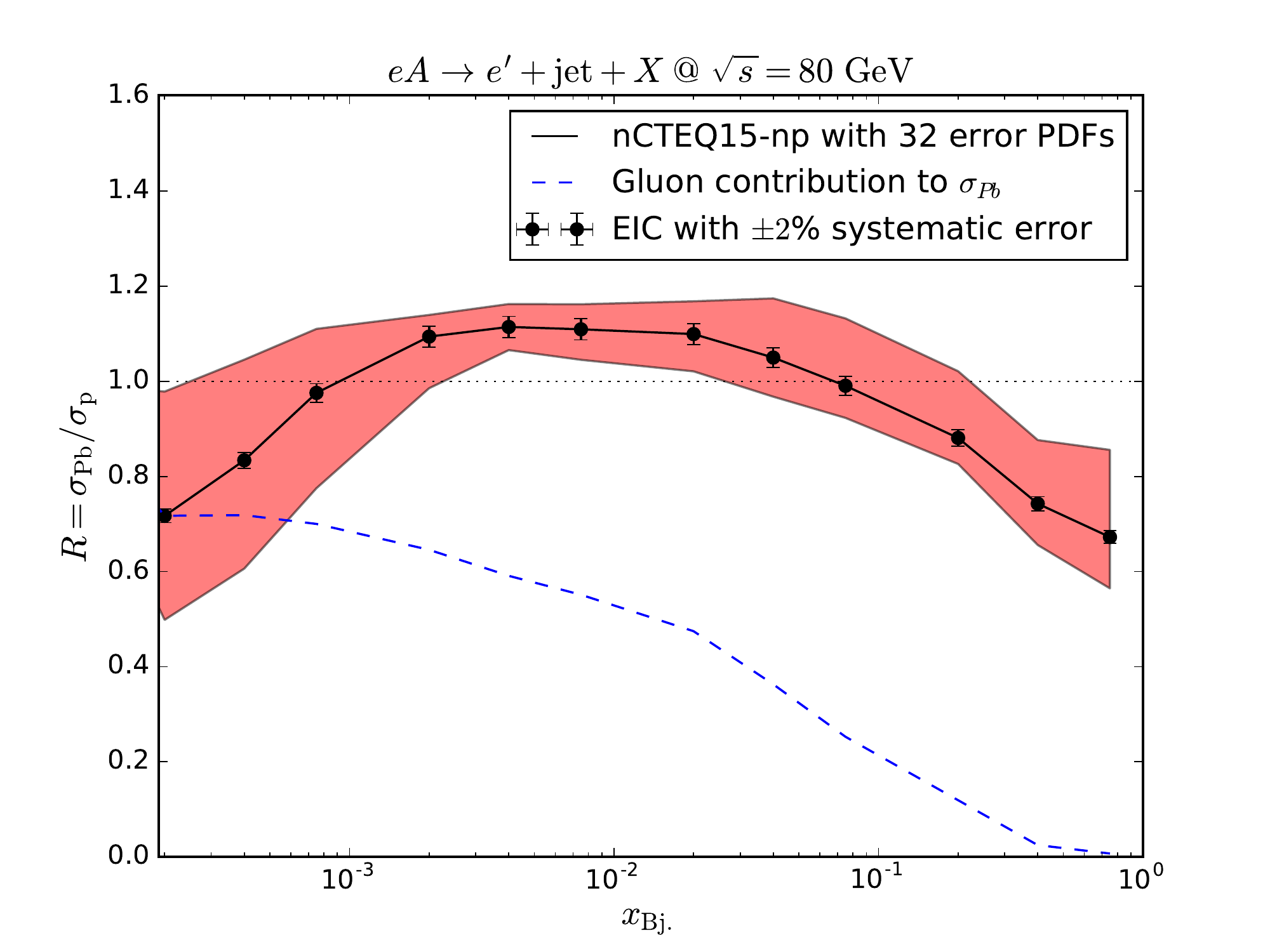,width=0.48\textwidth}
 \caption{\label{fig:4}Inclusive jet production in electron-lead ion collisions
 with beam energies of 16 and 100 GeV, respectively, at eRHIC. Shown is the
 ratio of electron-lead ion over electron-proton cross sections (full black lines)
 including the current nuclear PDF uncertainty from the nCTEQ15 fit to DIS and
 DY data only (red-shaded bands) as well as the relative gluon contribution to
 the total
 cross section (dashed blue lines) as a function of the jet transverse momentum
 (top left), rapidity (top right), photon virtuality (bottom left) and
 Bjorken-$x$ (bottom right). Error bars indicate the expected experimental
 precision.}
\end{figure}

The question is now what impact the EIC can have on a reduction of the
nuclear PDF uncertainties compared to our current knowledge from fixed-target
DIS and DY experiments. To this end, we show in Fig.\ \ref{fig:4} the same
central predictions as in the previous section of the nCTEQ15 fit to these
data for Pb-208 (full black lines), but supplement it now with the envelope of the
corresponding set of 32 error PDFs (red-shaded bands) determined with the
Hessian method \cite{Pumplin:2000vx,Pumplin:2001ct}. The latter relies on the
assumption that, near its minimum, the $\chi^2$-function can be approximated
by a quadratic form of the fitting parameters involving a matrix of
second-order partial derivatives with respect to the parameter shifts from
the minimum, which must then be diagonalized. We are particularly interested
in the gluon contribution, which suffers from the largest uncertainties
\cite{Kovarik:2015cma} and whose relative contribution to the differential
cross sections is therefore shown in addition (dashed blue lines).

What we observe at large rapidities (top right) and even more at small values
of $x$ (bottom right) is that the gluon contributes substantially there (up to
70\%) and that the nuclear PDF uncertainty reaches values of $\pm25$\%. In
these regions, the EIC would therefore have the greatest impact and might
eventually lead to a reduction of the uncertainty by an order of magnitude
(black error bars). A similar reduction of the gluon uncertainty has been
estimated to be possible in inclusive DIS and charm production at an EIC
with 20 GeV electrons and 100 GeV gold ions \cite{armesto} or at an LHeC
\cite{Helenius:2016hcu}.
The complementary regions, in particular the valence-quark dominated region at
medium-large $x$, have considerably smaller uncertainties of about $\pm10$\%,
which would, however, still be reduced with an EIC by a factor of five.
If one integrates over the rapidity, as has been done in the
$p_T$ (top left) and $Q^2$ (bottom left) distributions, the uncertainty in
the gluon-dominated regions at low values of $p_T$ and $Q^2$ shrinks
considerably, as one averages over large regions of $x$. At large $p_T$
and $Q^2$, however, one probes also large values of $x$, which can be
estimated by $x_T=2p_T/\sqrt{s}$ and $Q^2/(s\,y)$, respectively. At very
large $x$, information on nuclear PDFs is again very poor, as this region
is difficult to reach in fixed-target collisions, so that the nuclear PDF
uncertainty rises there to values of +30/$-$10\%.

\begin{figure}
 \epsfig{file=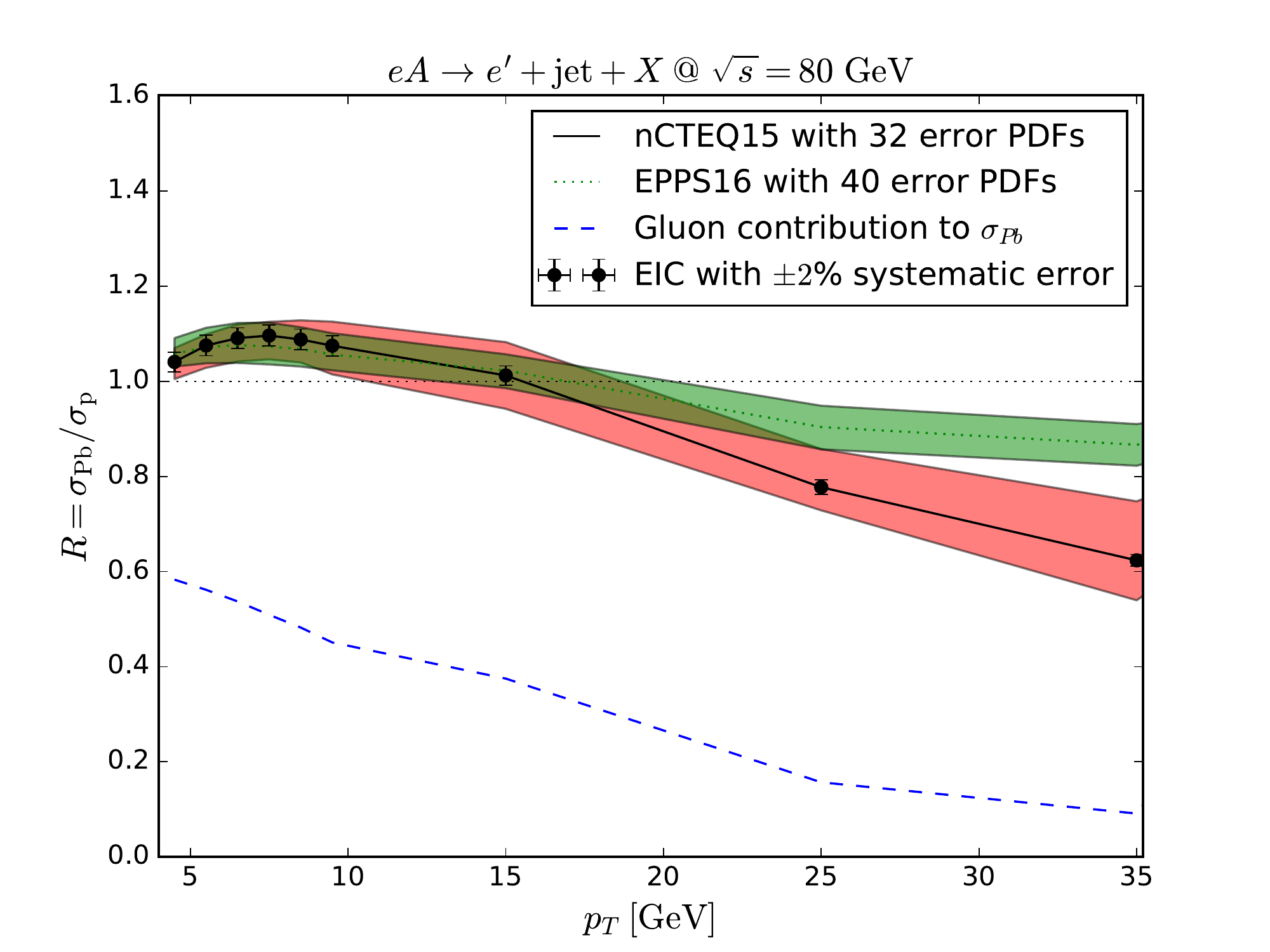,width=0.48\textwidth}
 \epsfig{file=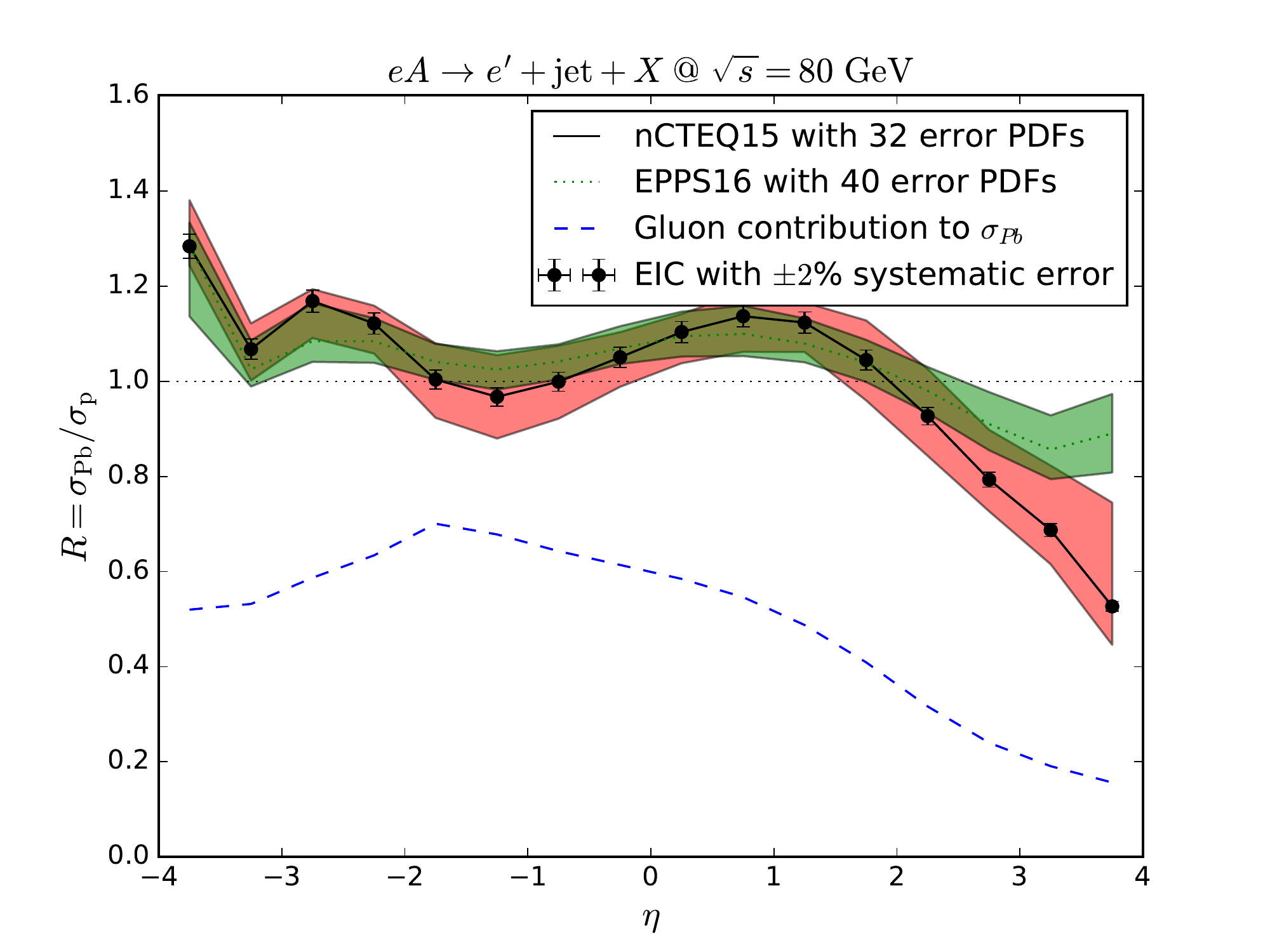,width=0.48\textwidth}
 \epsfig{file=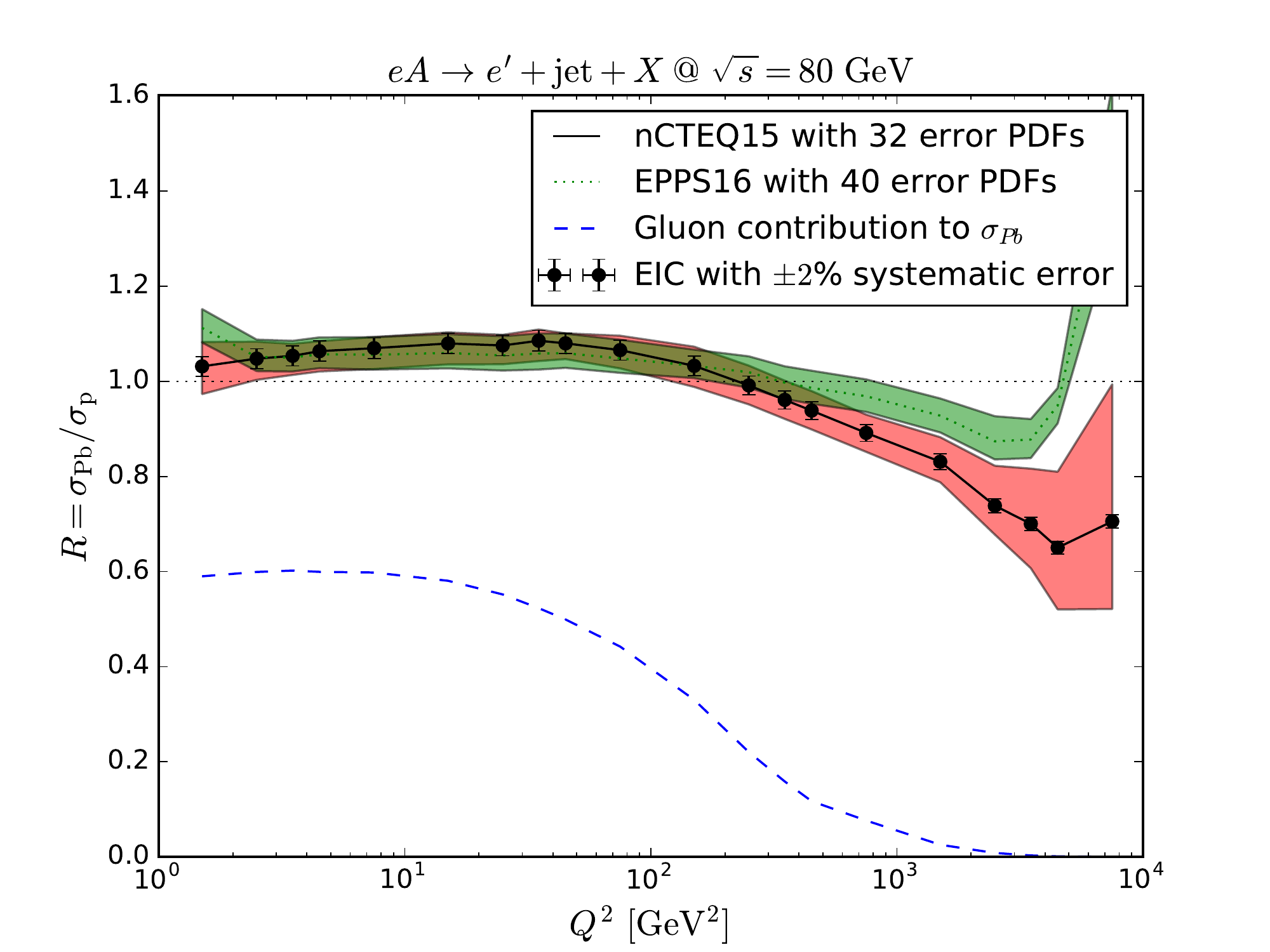,width=0.48\textwidth}
 \epsfig{file=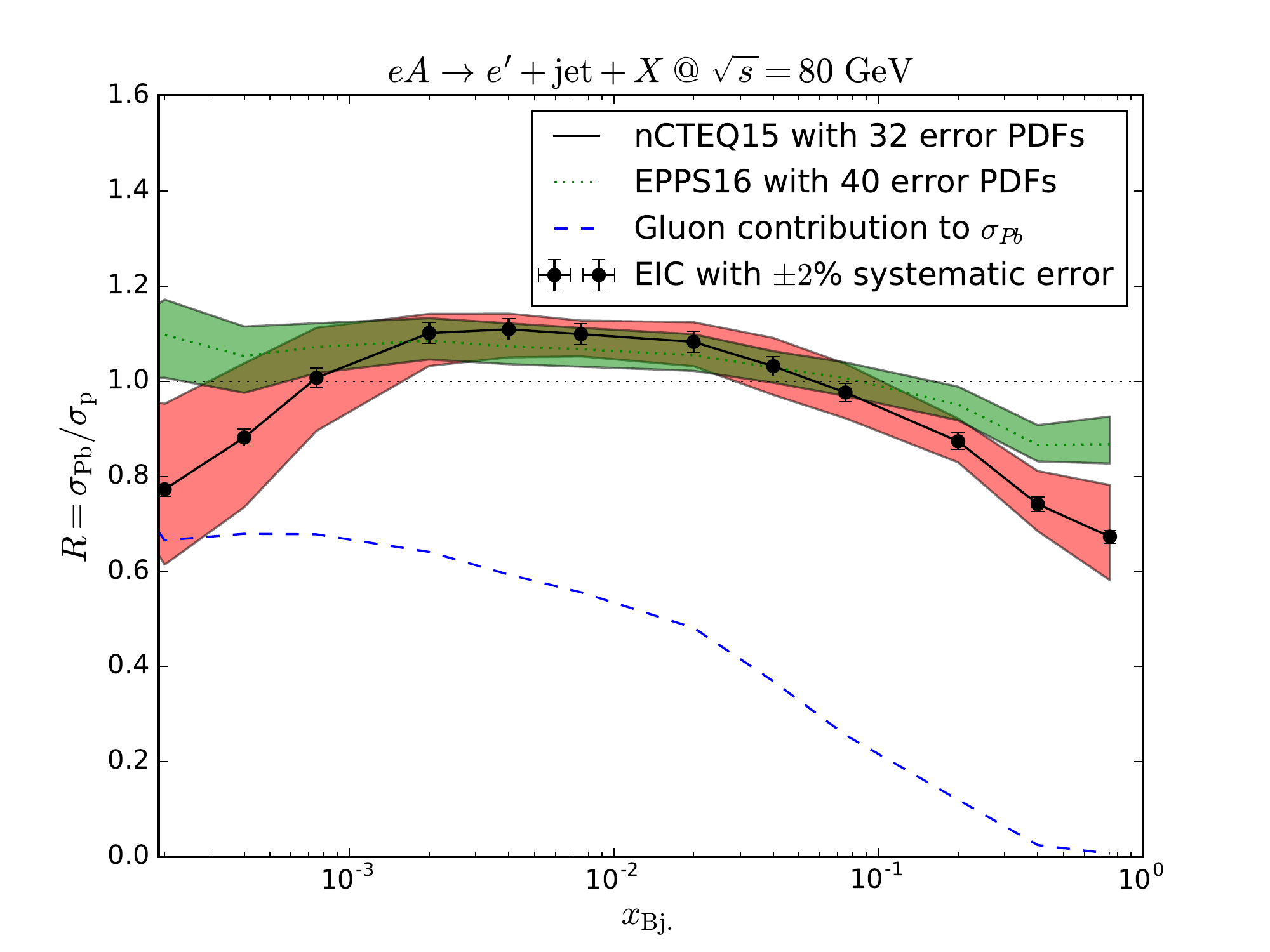,width=0.48\textwidth}
 \caption{\label{fig:5}Same as Fig.\ \ref{fig:4} for the nCTEQ15 fit including
 also inclusive pion data from D-Au collisions at BNL RHIC, and for the central
 EPPS16 fit (dotted green lines) to -- in particular -- dijet data from the LHC as
 well as the corresponding (green-shaded) error bands.}
\end{figure}

In our last figure, Fig.\ \ref{fig:5}, we repeat the same study as before,
but include now also inclusive pion data from BNL RHIC in the nCTEQ15 estimate
of the nuclear PDF uncertainty. As we mentioned before, this additional
information depends on theoretical assumptions about the fragmentation
function of quarks and gluons into pions. Comparing Figs.\ \ref{fig:4} and
\ref{fig:5} reveals that this additional information reduces the uncertainty
by about a third, in particular at large rapidity and low Bjorken-$x$,
but also at large $p_T$ and $Q^2$, but that even under this additional
assumption there is still large room for improvement from the EIC (black
error bars).

It is interesting to confront the full nCTEQ15 fit using the inclusive
pion data from D-Au collisions at BNL RHIC (full black lines and red-shaded
bands) with another, even more recent nPDF
analysis, EPPS16 \cite{Eskola:2009uj}, which also includes these data, but in addition uses
CERN LHC data on $W$ and $Z$ production and, more importantly, dijet
production in $p$-Pb collisions at a center-of-mass energy of 5.02 TeV
\cite{Chatrchyan:2014hqa}. Therefore, in Fig.\ \ref{fig:5} the central EPPS16
predictions are also shown (dotted green lines), together with the envelope of
the corresponding set of 40 error PDFs (green-shaded bands), determined again
with the Hessian method. Overall, one observes that the shapes of the cross
section ratios in the four distributions differ somewhat, in particular
at the kinematic edges. While at low $p_T$ (top left) nCTEQ15 and EPPS16 make
very similar predictions, at high $p_T$ EPPS16 predict about half the
suppresion from
nCTEQ15 with an uncertainty that is also about half as big. This is not
surprising, as the fitted CMS dijet production data extend to jets of
$p_T\simeq40$ GeV \cite{Chatrchyan:2014hqa}, which is much higher than the
$p_T<16$ and 17 GeV pions that were measured with PHENIX \cite{Adler:2006wg}
and STAR \cite{Abelev:2009hx} at BNL RHIC, respectively. Similarly, EPPS16
fitted to CMS dijet data with rapidities up to $\eta<2.5$, while the pion
measurements by PHENIX and STAR extended only to $|\eta|\leq0.35$ and $0<
\eta<1$, respectively, so that differences at very forward rapidities are
to be expected (top right). The reduced uncertainty there translates into
a similarly reduced uncertainty at
low Bjorken-$x$ (bottom right), while in the $Q^2$ distribution the nCTEQ15
and EPPS16 predictions are again very similar, except at very high scales
(bottom left). Under the assumption that jets are not (or at least less than
pions) modified in $pA$ collisions, the EPPS16 predictions are already quite
precise, but would still be improved at an EIC by a factor of up to five.


\section{Conclusion and outlook}
\label{sec:5}

Let us therefore now come to our conclusions. In this paper, we have made
predictions for inclusive jet production in electron-ion collisions at a
possible future EIC. Our goal was in particular to establish the benefit
that such a collider might have on a more precise determination of nuclear
PDFs, which is not only required to enhance our knowledge of quark and gluon
dynamics in the nucleus, but also to allow for a reliable extraction of
hot nuclear matter properties after a proper subtraction of cold nuclear
effects.
Theoretically, our calculations were based on a full NLO and an approximate
NNLO calculation, implemented in the program {\tt JetViP}. While the NLO
corrections were large, in particular at low perturbative scales, perturbative
stability was restored at aNNLO in line with expectations from full NNLO
calculations.
Phenomenologically, we have established that measurements of inclusive jet
production at an EIC would extend the kinematic ranges to $Q^2\leq10^3$
GeV$^2$ and $x\geq10^{-4}$ similarly to inclusive DIS and allow to reduce
the uncertainty on nuclear PDFs, in particular the one of the gluon at low
$x$, by factors of five to ten. This improvement would probably not be 
possible in inclusive DIS alone, but would alternatively require additional
charm tagging possibilities.

Future calculations could properly include jet mass effects in the aNNLO
calculation \cite{deFlorian:2013qia}, although as we have seen the impact
of these corrections is small, and extend the present study to dijet
production, which would allow for more complete kinematic constraints.
More differential studies of single, two and three jets and their shapes
at the EIC might help to establish if they are modified in $eA$ collisions
compared to $ep$ collisions,
similarly to the modification of the pion fragmentation function in $AA$
collisions and possible collective effects in $pA$ collisions. It would
then become possible to investigate transport properties of the cold nuclear
medium and test the strong gluon field paradigm \cite{vitev}.
Finally, even transverse-momentum dependent distribution functions (TMDs)
of gluons in protons and nuclei might become accessible in measurements of
dijet asymmetries in polarized or unpolarized $ep$ and $eA$ collisions
\cite{Boer:2016fqd}.


\acknowledgments

We thank the organizers of the 7th International Conference on {\it Physics
Opportunities at an ElecTron-Ion-Collider} (POETIC 7), which motivated
this study, for the kind invitation and C.\ Klein-B\"osing for useful
discussions. This work has been supported by the BMBF under contract
05H15PMCCA. All figures have been produced using {\tt Matplotlib}
\cite{Hunter:2007}.


\bibliographystyle{apsrev}

\begin{thebibliography}{00}

\bibitem{Collins:1989gx} 
  J.~C.~Collins, D.~E.~Soper and G.~F.~Sterman,
  Adv.\ Ser.\ Direct.\ High Energy Phys.\  {\bf 5}, 1 (1989).

\bibitem{Butterworth:2015oua} 
  J.~Butterworth {\it et al.},
  J.\ Phys.\ G {\bf 43}, 023001 (2016).

\bibitem{Accardi:2012qut} 
  A.~Accardi {\it et al.},
  Eur.\ Phys.\ J.\ A {\bf 52}, 268 (2016).

\bibitem{armesto}
 N.~Armesto,
 {\it Nuclear PDFs at an EIC},
 talk given at the 7th International Conference on {\it Physics Opportunities
 at an ElecTron-Ion-Collider} (POETIC 7),
 Philadelphia, Nov.\ 14, 2016.

\bibitem{deFlorian:2011fp} 
  D.~de Florian, R.~Sassot, P.~Zurita and M.~Stratmann,
  Phys.\ Rev.\ D {\bf 85}, 074028 (2012).

\bibitem{Kovarik:2015cma} 
  K.~Kovarik {\it et al.},
  Phys.\ Rev.\ D {\bf 93}, 085037 (2016).

\bibitem{Eskola:2009uj} 
  K.~J.~Eskola, H.~Paukkunen and C.~A.~Salgado,
  JHEP {\bf 0904}, 065 (2009);
%
  K.~J.~Eskola, P.~Paakkinen, H.~Paukkunen and C.~A.~Salgado,
  arXiv:1612.05741 [hep-ph].

\bibitem{Armesto:2015lrg} 
  N.~Armesto, H.~Paukkunen, J.~M.~Penín, C.~A.~Salgado and P.~Zurita,
  Eur.\ Phys.\ J.\ C {\bf 76}, 218 (2016).

\bibitem{Hirai:2007sx} 
  M.~Hirai, S.~Kumano and T.-H.~Nagai,
  Phys.\ Rev.\ C {\bf 76}, 065207 (2007);
%
  M.~Hirai,
  JPS Conf.\ Proc.\  {\bf 12}, 010024 (2016).

\bibitem{Schienbein:2009kk} 
  I.~Schienbein, J.~Y.~Yu, K.~Kovarik, C.~Keppel, J.~G.~Morfin, F.~Olness and J.~F.~Owens,
  Phys.\ Rev.\ D {\bf 80}, 094004 (2009).

\bibitem{Brandt:2013hoa} 
  M.~Klasen and M.~Brandt,
  Phys.\ Rev.\ D {\bf 88}, 054002 (2013);
%
  M.~Brandt, M.~Klasen and F.~K\"onig,
  Nucl.\ Phys.\ A {\bf 927}, 78 (2014).

\bibitem{Kusina:2016fxy} 
  A.~Kusina {\it et al.},
  arXiv:1610.02925 [nucl-th].

\bibitem{Abelev:2012ola} 
  B.~Abelev {\it et al.} [ALICE Collaboration],
  Phys.\ Lett.\ B {\bf 719}, 29 (2013);
%
  B.~B.~Abelev {\it et al.} [ALICE Collaboration],
  Phys.\ Lett.\ B {\bf 726}, 164 (2013);
%
  B.~B.~Abelev {\it et al.} [ALICE Collaboration],
  Phys.\ Rev.\ C {\bf 90}, 054901 (2014).

\bibitem{Klasen:1995ab} 
  M.~Klasen and G.~Kramer,
  Z.\ Phys.\ C {\bf 72}, 107 (1996);
%
  M.~Klasen and G.~Kramer,
  Z.\ Phys.\ C {\bf 76}, 67 (1997);
%
  M.~Klasen, T.~Kleinwort and G.~Kramer,
  Eur.\ Phys.\ J.\ direct C {\bf 1}, 1 (1998).

\bibitem{Klasen:1997jm} 
  M.~Klasen, G.~Kramer and B.~P\"otter,
  Eur.\ Phys.\ J.\ C {\bf 1}, 261 (1998).

\bibitem{Klasen:2013cba} 
  M.~Klasen, G.~Kramer and M.~Michael,
  Phys.\ Rev.\ D {\bf 89}, 074032 (2014).

\bibitem{Biekotter:2015nra} 
  T.~Biek\"otter, M.~Klasen and G.~Kramer,
  Phys.\ Rev.\ D {\bf 92}, 074037 (2015).

\bibitem{Kidonakis:2003tx} 
  N.~Kidonakis,
  Int.\ J.\ Mod.\ Phys.\ A {\bf 19}, 1793 (2004).

\bibitem{Abelof:2016pby} 
  G.~Abelof, R.~Boughezal, X.~Liu and F.~Petriello,
  Phys.\ Lett.\ B {\bf 763}, 52 (2016).

\bibitem{Currie:2016ytq} 
  J.~Currie, T.~Gehrmann and J.~Niehues,
  Phys.\ Rev.\ Lett.\  {\bf 117}, 042001 (2016).

\bibitem{yoshida}
 R.~Yoshida,
 {\it Jefferson Lab EIC: Physics, accelerator and detector},
 talk given at the 7th International Conference on {\it Physics Opportunities
 at an ElecTron-Ion-Collider} (POETIC 7),
 Philadelphia, Nov.\ 15, 2016.

\bibitem{mueller}
 B.~M\"uller,
 {\it eRHIC - an EIC at BNL},
 talk given at the 7th International Conference on {\it Physics Opportunities
 at an ElecTron-Ion-Collider} (POETIC 7),
 Philadelphia, Nov.\ 15, 2016.

\bibitem{Potter:1999gg} 
  B.~P\"otter,
  Comput.\ Phys.\ Commun.\  {\bf 133}, 105 (2000).

\bibitem{Andreev:2014wwa} 
  V.~Andreev {\it et al.} [H1 Collaboration],
  Eur.\ Phys.\ J.\ C {\bf 75}, 65 (2015).

\bibitem{Cacciari:2008gp} 
  M.~Cacciari, G.~P.~Salam and G.~Soyez,
  JHEP {\bf 0804}, 063 (2008).

\bibitem{Ellis:1993tq} 
  S.~D.~Ellis and D.~E.~Soper,
  Phys.\ Rev.\ D {\bf 48}, 3160 (1993).

\bibitem{Armesto:2006ph} 
  N.~Armesto,
  J.\ Phys.\ G {\bf 32}, R367 (2006).

\bibitem{Frankfurt:2003zd} 
  L.~Frankfurt, V.~Guzey and M.~Strikman,
  Phys.\ Rev.\ D {\bf 71}, 054001 (2005).

\bibitem{Baltz:2007kq} 
  A.~J.~Baltz {\it et al.},
  Phys.\ Rept.\  {\bf 458}, 1 (2008).

\bibitem{Guzey:2016tek} 
  V.~Guzey and M.~Klasen,
  JHEP {\bf 1604}, 158 (2016).

\bibitem{Arneodo:1992wf} 
  M.~Arneodo,
  Phys.\ Rept.\  {\bf 240}, 301 (1994).

\bibitem{Geesaman:1995yd} 
  D.~F.~Geesaman, K.~Saito and A.~W.~Thomas,
  Ann.\ Rev.\ Nucl.\ Part.\ Sci.\  {\bf 45}, 337 (1995).

\bibitem{Brodsky:2004qa} 
  S.~J.~Brodsky, I.~Schmidt and J.~J.~Yang,
  Phys.\ Rev.\ D {\bf 70}, 116003 (2004).

\bibitem{Frankfurt:2016qca} 
  L.~Frankfurt, V.~Guzey and M.~Strikman,
  arXiv:1612.08273 [hep-ph].

\bibitem{Pumplin:2000vx} 
  J.~Pumplin, D.~R.~Stump and W.~K.~Tung,
  Phys.\ Rev.\ D {\bf 65}, 014011 (2001).

\bibitem{Pumplin:2001ct} 
  J.~Pumplin, D.~Stump, R.~Brock, D.~Casey, J.~Huston, J.~Kalk, H.~L.~Lai and W.~K.~Tung,
  Phys.\ Rev.\ D {\bf 65}, 014013 (2001).

\bibitem{Helenius:2016hcu} 
  I.~Helenius, H.~Paukkunen and N.~Armesto,
  PoS DIS {\bf 2016}, 276 (2016)
  [arXiv:1606.09003 [hep-ph]].

\bibitem{Chatrchyan:2014hqa} 
  S.~Chatrchyan {\it et al.} [CMS Collaboration],
  Eur.\ Phys.\ J.\ C {\bf 74}, 2951 (2014).

\bibitem{Adler:2006wg} 
  S.~S.~Adler {\it et al.} [PHENIX Collaboration],
  Phys.\ Rev.\ Lett.\  {\bf 98}, 172302 (2007).

\bibitem{Abelev:2009hx} 
  B.~I.~Abelev {\it et al.} [STAR Collaboration],
  Phys.\ Rev.\ C {\bf 81}, 064904 (2010).

\bibitem{deFlorian:2013qia} 
  D.~de Florian, P.~Hinderer, A.~Mukherjee, F.~Ringer and W.~Vogelsang,
  Phys.\ Rev.\ Lett.\  {\bf 112}, 082001 (2014).

\bibitem{vitev}
 I.~Vitev,
 {\it Opportunities with jet physics in e+A collisions},
 talk given at the 7th International Conference on {\it Physics Opportunities
 at an ElecTron-Ion-Collider} (POETIC 7),
 Philadelphia, Nov.\ 15, 2016.

\bibitem{Boer:2016fqd} 
  D.~Boer, P.~J.~Mulders, C.~Pisano and J.~Zhou,
  JHEP {\bf 1608}, 001 (2016).

\bibitem{Hunter:2007}
  J.~D.~Hunter, 
  Computing In Science \& Engineering {\bf 9}, 90 (2007).
 
\end{thebibliography}


\end{document}